\documentclass[journal, twocolumn, 10pt]{IEEEtran}
\IEEEoverridecommandlockouts
\usepackage{cite}
\usepackage{graphicx}
\usepackage{amsfonts}
\usepackage{amsmath}
\usepackage{physics}
\usepackage{amsthm}
\usepackage{amssymb}
\usepackage{verbatim}
\usepackage{subcaption}
\usepackage{hyperref}
\usepackage{color}
\usepackage{bbm}
\usepackage{babel,blindtext}
\usepackage[utf8]{inputenc}

\DeclareMathOperator*{\argmax}{arg\,max}
\DeclareMathOperator*{\argmin}{arg\,min}

\newtheorem{theorem}{Theorem}

\newtheorem{lemma}[theorem]{Lemma}
\newtheorem{proposition}[theorem]{Proposition}
\newtheorem{remark}{Remark}
\newtheorem{example}{Example}
\newtheorem{definition}{Definition}

\title{Optimal Private Discrete Distribution Estimation with One-bit Communication}

\author{{Seung-Hyun~Nam,~\IEEEmembership{Graduate Student Member,~IEEE,} Vincent Y. F. Tan,~\IEEEmembership{Senior Member,~IEEE,} and \\ Si-Hyeon Lee,~\IEEEmembership{Senior Member,~IEEE.}}
\thanks{Seung-Hyun Nam and Si-Hyeon Lee are with the School of Electrical Engineering, Korea Advanced Institute of Science and Technology (KAIST), Daejeon 34141, South Korea (e-mail: shnam@kaist.ac.kr; sihyeon@kaist.ac.kr).
Vincent Y. F. Tan is with the Department of Mathematics, National
University of Singapore, Singapore 119076, and also with the Department
of Electrical and Computer Engineering, National University of Singapore,
Singapore 117583 (e-mail: vtan@nus.edu.sg). (Corresponding author: Si-Hyeon Lee)

This article has supplementary material provided by the authors.
}
}

\begin{document}

\maketitle

\begin{abstract}
    We consider a private discrete distribution estimation problem with one-bit communication constraint.
    The privacy constraints are imposed with respect to the local differential privacy and the maximal leakage.
    The estimation error is quantified by the worst-case mean squared error.
    We completely characterize the first-order asymptotics of this privacy-utility trade-off under the one-bit communication constraint for both types of privacy constraints by using ideas from local asymptotic normality and the resolution of a block design mechanism.
    These results demonstrate the optimal dependence of the privacy-utility trade-off under the one-bit communication constraint in terms of the parameters of the privacy constraint and the size of the alphabet of the discrete distribution.
\end{abstract}

\begin{IEEEkeywords}
Discrete distribution estimation, local differential privacy, maximal leakage,  one-bit communication, privacy-utility-communication trade-off.
\end{IEEEkeywords}

\section{Introduction}
Statistical inference problems under privacy constraints have been studied extensively in recent years~\cite{Kasivis11_PrivLearn, Calmon12_Priv_Stat_inf, duchi14_LDP_minimax, Sankar13_PUT, Dwork14_found_dp, kairouz14_extremal, Kairouz16_DDE_LDP, Bhowmick18_PrivUnit, Ye18_SS, Asoodeh19_Priv_Est_Eff, issa19_operational, Liao19_alphaleakage, HR-19Acharya, barnes20_fisher_LDP, geng20_ADP_PUT, PGR-22Feldman, park23_block, ye17_opt_PUT_l2}.
Among numerous well-established privacy metrics, {\em local differential privacy} (LDP) has emerged as one of the most popular privacy requirements \cite{Kasivis11_PrivLearn, duchi14_LDP_minimax, kairouz14_extremal}.
The LDP restricts the amount of leakage of private information from the released data of individuals.
It also admits an operational definition in terms of the fundamental limits of the probability of adversarial guess \cite[Thm.~14]{issa19_operational}.
Together with the LDP, the {\em maximal leakage} (ML) also limits the amount of leakage of private information.
%It also admits an operational definition.
In contrast to the LDP taking into account the worst-case leakage, the ML considers the average leakage \cite[Thm.~1]{issa19_operational}.
In a private statistical inference problem, there is a fundamental trade-off between the amount of privacy leakage and the inference error as data should be perturbed before released to satisfy the privacy constraint.
This is known as the {\em privacy-utility trade-off} (PUT).
The PUTs for various private inference problems have been studied \cite{Calmon12_Priv_Stat_inf, duchi14_LDP_minimax, Sankar13_PUT, Dwork14_found_dp, kairouz14_extremal, Kairouz16_DDE_LDP, Bhowmick18_PrivUnit, Ye18_SS, Asoodeh19_Priv_Est_Eff, issa19_operational, Liao19_alphaleakage, HR-19Acharya, barnes20_fisher_LDP, geng20_ADP_PUT, PGR-22Feldman, park23_block, ye17_opt_PUT_l2}.
In particular, Ye and Barg~\cite{ye17_opt_PUT_l2} completely characterized the optimal PUT for discrete distribution estimation under the LDP constraint.

% Apart from the privacy constraint, communication constraints have been also considered in \cite{Ahlswede86_comm_HT, Longo_comm_HT_decent, Xiang12_Int_Comm_HT, Zhang13_comm_est_LB, Han18_Geo_comm_est_LB, barnes20_Comm_Fisher, Acharya20_comm_SR}.
% Those works seek to find communication efficient inference schemes.
In addition to privacy, another important factor of practical interest is the {\em communication cost} to send the individual's data.
%In this paper, we take into account both privacy and communication constraints.
It is rather natural that there exists a fundamental trade-off between the amount of privacy leakage, the quality of inference, and the communication cost.
We coin this as the {\em privacy-utility-communication trade-off} (PUCT).
The PUCTs for different types of inference problems have been studied \cite{acharya19_PUCT, pensia23_HT_PUCT, acharya21_Sparse_est_PUCT, chen20_trilemma, nam22_onebit_conv}.
In particular, \cite{chen20_trilemma} characterized the PUCTs for mean estimation, frequency estimation, and discrete distribution estimation in the order-optimal sense, which means that the upper and lower bounds may differ up to some constants.
These results, while useful, might be far from the optimal PUCT because the underlying multiplicative constant factors are not quantified.
Also, \cite{nam22_onebit_conv} analyzed the optimal PUCT up to the factor of $4$ for discrete distribution estimation with the minimum communication cost, i.e., the one-bit communication constraint.

In this paper, we consider  the private discrete distribution estimation problem, %, which is one of the most fundamental statistical inference problems.
with two privacy constraints, namely, the LDP constraint and the ML constraint.
As the most communication-cost effective setting,  we consider the one-bit communication constraint which  allows the minimum non-trivial amount of communication. 
The estimation error is set to be the worst-case mean squared error (MSE).
Our main result for this setup is rather simple but conclusive:
we completely characterize the first-order asymptotics of the PUT under the one-bit communication constraint for both the LDP constraint and the ML constraint, where the asymptotics is in the number of clients $n$.
To do so, we prove impossibility results and propose optimal schemes based on novel block design mechanisms \cite{park23_block,nam23_res_BD}. 

\subsection{Related works}
The literature on statistical inference under privacy and/or communication constraints is vast.
Among them, we introduce the works which consider discrete distribution estimation under the LDP or the ML as the privacy constraint, and MSE as the error of the estimation.
Duchi {\em et al.}~\cite{duchi14_LDP_minimax} established the minimax framework on private parametric estimation and provided an order-optimal PUT under the $\epsilon$-LDP constraint for $\epsilon \in (0,1]$.
Also, the authors proposed a method to derive a lower bound of the PUT based on Le Cam's, Fano's, and Assouad's methods and a \emph{strong data processing inequality}.
Later, Ye and Barg~\cite{Ye18_SS} proposed the \emph{subset selection} scheme and this was shown to achieve the optimal PUT under the $\epsilon$-LDP constraint for all $\epsilon>0$~\cite{ye17_opt_PUT_l2}.
A tight lower bound of PUT was derived by using the concept of \emph{local asymptotic normality}~\cite{le00_asymptotics,ibragimov13_stat_asymp,van98_asymptotic}.

Concerning the PUCT, Chen {\em et al.}~\cite{chen20_trilemma} analyzed an order-optimal PUCT under the $\epsilon$-LDP and the $b$-bit communication constraints for all $\epsilon>0$ and $b \geq 1$.
The authors proposed the \emph{recursive Hadamard response} as an achievability scheme, and an order-optimal lower bound was derived by combining the lower bounds from Ye and Barg~\cite{Ye18_SS} (for the LDP constraint), and Barnes {\em et al.}~\cite{barnes20_Comm_Fisher} (for the communication constraint).
The lower bound in~\cite{barnes20_Comm_Fisher} was derived by deriving an upper bound of the trace of the Fisher information matrix and applying the \emph{van Trees inequality}~\cite{gill95_VTineq}.
These techniques were also modified to derive a lower bound of PUT~\cite{barnes20_fisher_LDP}.
Under the one-bit communication constraint, Nam and Lee~\cite{nam22_onebit_conv} proposed a tighter lower bound which meets the upper bound achieved by the recursive Hadamard response up to the factor of $4$.
The lower bound in~\cite{nam22_onebit_conv} was derived by modifying the van Trees inequality into a \emph{symmetric} version, and maximizing the trace of the Fisher information matrix by exploiting the extreme points of the set of $\epsilon$-LDP mechanisms with one-bit output.
The extreme points of the set of $\epsilon$-LDP mechanisms were studied by Holohan \emph{et al.}~\cite{holohan17_LDP_ex_pts}, and a similar idea was considered by Kairouz \emph{et al.}~\cite{kairouz14_extremal}.
For the upper bound of the PUCT, Park {\em et al.}~\cite{park23_block} proposed a class of {\em block design schemes} which achieve the optimal PUT with low communication costs.
This class subsumes many previous schemes such as the {\em subset selection} by Ye and Barg~\cite{Ye18_SS}, the {\em Hadamard response} by Acharya {\em et al.}~\cite{HR-19Acharya}, and the {\em projective geometry response} by Feldman {\em et al.}~\cite{PGR-22Feldman}.
Recently, Nam {\em et al.}~\cite{nam23_res_BD} proposed a method to reduce the communication cost of a block design scheme by exploiting shared randomness.
The authors showed that one-bit of communication is sufficient to achieve the optimal PUT under the $\epsilon$-LDP constraint for all  $\epsilon \leq \frac{1}{2}\log \frac{v+2}{v-2}$ and even $v$, where $v$ denotes the size of the alphabet of the discrete distribution.  
%the scheme so-called {\em Baranyai's resolution of the subset selection} that uses a one-bit of communication achieves the optimal PUCT under the $\epsilon$-LDP constraint and $b$-bit communication constraint for all $b\geq 1$, $\epsilon \leq \frac{1}{2}\log \frac{v+2}{v-2}$, and $v$ is even, where $v$ denotes the alphabet size of the discrete distribution. 

In this work, we extend the above contributions by proposing a unifying framework to derive the {\em exact} first-order asymptotics of the PUT under either of the $(\epsilon,\delta)$-LDP and the $\gamma$-ML privacy constraints as well as the one-bit communication constraint.

\subsection{Paper outline}
The rest of this paper is organized as follows.
In Section~\ref{sec:model}, we formulate the problem of private discrete distribution estimation under the one-bit communication constraint.
In Section~\ref{sec:result}, we present the main theorem that characterizes the PUTs and briefly discuss the ideas behind the proofs, which are related to the model with shared randomness.
Accordingly, we present the model with shared randomness in Section~\ref{sec:model_SR}.
In Sections~\ref{sec:conv} and~\ref{sec:achiev}, we prove the converse (lower bounds on PUT) and the achievability (upper bounds on PUT) parts of the proof of the main theorem, respectively.
Finally, Section~\ref{sec:conc} concludes the paper.

\subsection{Notations}
For integers $a<b$, we denote $[a:b] := \{a,a+1,\ldots,b\}$, and we write $[a] := [1:a]$.
For a finite set $\mathcal{X}$, $x^n \in \mathcal{X}^n$, and $\mathcal{I} = (i_1,\ldots,i_t) \in [n]^t$, $x_\mathcal{I}$ denotes  $(x_{i_1},\ldots,x_{i_t})$. 
We write $\mathbf{0}$ as the all-zeros vector, $\mathbf{1}$ as an all-ones vector or matrix, and $I$ as the identity matrix of a suitable dimension which will be clear from the context.
If these quantities are indexed by a subscript, the subscript denotes the dimension.
For finite sets $\mathcal{X}$ and $\mathcal{Y}$, we denote
$\mathcal{P}(\mathcal{X})$ as the set of all probability mass functions on $\mathcal{X}$, and a conditional probability mass function $Q$ from $\mathcal{X}$ to $\mathcal{Y}$ as $Q:\mathcal{X} \rightarrow \mathcal{P}(\mathcal{Y})$.
We say that two conditional probability mass functions $Q_1:\mathcal{X} \rightarrow \mathcal{P}(\mathcal{Y})$ and $Q_2:\mathcal{Z} \rightarrow \mathcal{P}(\mathcal{W})$ are \emph{equivalent} if 
\begin{equation}
    \forall x\in \mathcal{X}, y \in \mathcal{Y}, \quad Q_1(y|x) = Q_2(\psi_2(y)|\psi_1(x)),
\end{equation}
for some bijections $\psi_1:\mathcal{X} \rightarrow \mathcal{Z}$ and $\psi_2 : \mathcal{Y} \rightarrow \mathcal{W}$, or more succinctly, $Q_1 \cong Q_2$.
For a conditional probability mass function $Q:\mathcal{X}\rightarrow \mathcal{P}(\mathcal{Y})$, we will also treat $Q$ as a (row) stochastic matrix whose row and column indices correspond to $\mathcal{X}$ and $\mathcal{Y}$, respectively.

\section{System Model}\label{sec:model}

\begin{figure}
    \centering
    \includegraphics[width=0.9\linewidth]{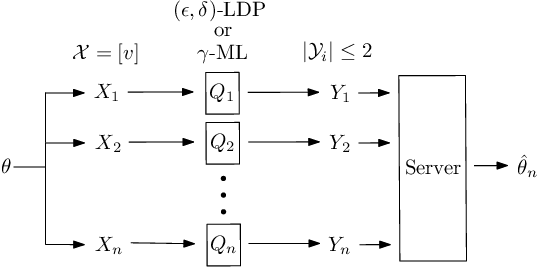}
    \caption{Discrete distribution estimation under a privacy constraint and a one-bit communication constraint.}
    \label{fig:model}
\end{figure}

We consider discrete distribution estimation under two constraints, a privacy constraint and a one-bit communication constraint.
The setup is depicted in Fig.~\ref{fig:model}.
In this model, there are $n$ clients.
The $i$-th client has its own data $X_i \in \mathcal{X} = [v]$ where the alphabet size $v \in \mathbb{Z}_{\geq 2}$.
We assume that $X_1,\ldots,X_n$ are i.i.d.\ random variables with $X_i \sim \theta$, where $\theta \in \mathcal{P}([v])$ is an unknown probability mass function supported on $[v]$.
To prevent leakage of private information, each of the $n$ clients randomly perturbs its data $X_i$ into $Y_i$ through a conditional probability mass function $Q_i:\mathcal{X} \rightarrow \mathcal{P}(\mathcal{Y}_i)$, which we call a \emph{privacy mechanism}.
Without loss of generality, we assume that for all $y \in \mathcal{Y}_i$, $Q_i(y|x) > 0$ for some $x \in \mathcal{X}$.
In this work, we consider two types of privacy constraints, namely, the  $(\epsilon,\delta)$-\textbf{local differential privacy} and the $\gamma$-\textbf{maximal leakage} constraints \cite{kairouz14_extremal, issa19_operational}.
\begin{definition}\label{def:priv}
    For $\epsilon > 0$ and $\delta \in [0,1]$, a privacy mechanism $Q:\mathcal{X} \rightarrow \mathcal{P}(\mathcal{Y})$ is said to be an {\em $(\epsilon,\delta)$-local differential privacy (LDP) mechanism} if
    \begin{equation}
        \forall y\in \mathcal{Y},\; x,x' \in \mathcal{X}, \quad Q(y|x) \leq e^\epsilon Q(y|x') + \delta.
    \end{equation}
    For $\gamma > 0$, a privacy mechanism $Q:\mathcal{X} \rightarrow \mathcal{P}(\mathcal{Y})$ is said to be a {\em $\gamma$-maximal leakage (ML) mechanism} if
    \begin{equation}
        \sum_{y\in\mathcal{Y}} \max_{x\in \mathcal{X}} Q(y|x) \leq e^\gamma.
    \end{equation}
\end{definition}
Together with the privacy constraint, we also consider the one-bit communication constraint to minimize the amount of communication.
A privacy mechanism $Q:\mathcal{X} \rightarrow \mathcal{P}(\mathcal{Y})$ is said to satisfy the \textbf{one-bit communication constraint} if $|\mathcal{Y}| \leq 2$.
Under the one-bit communication constraint, the $\gamma$-ML constraint becomes vacuous when $\gamma > \log 2$.
Thus, we will only consider $\gamma \leq \log 2$.
For notational simplicity, we define $\mathcal{Q}^{(\epsilon,\delta)}$ as the set of all $(\epsilon,\delta)$-LDP mechanisms satisfying the one-bit communication constraint, and $\mathcal{Q}^{\gamma}$ as the set of all $\gamma$-ML mechanisms satisfying the one-bit communication constraint.
Also, we will simply write $\mathcal{Q}$ as either $\mathcal{Q}^{(\epsilon,\delta)}$ or $\mathcal{Q}^{\gamma}$ for statements that do not depend on the choice of the privacy constraint.
Then, the constraints on the privacy mechanisms $Q_1, \ldots, Q_n$ can be simply written as
\begin{equation}
    \forall i \in [n], \quad Q_i \in \mathcal{Q}. \label{eq:const_simple}
\end{equation}

After the clients perturb their data to $Y^n$, the server collects them and estimates the unknown distribution of data $\theta$ using an {\em estimator} $\hat{\theta}_n : \mathcal{Y}^n \rightarrow \mathbb{R}^v$.
We call a tuple of privacy mechanisms satisfying the constraint~\eqref{eq:const_simple} and an estimator $\hat{\theta}_n$, $(Q_1,\ldots,Q_n,\hat{\theta}_n)$ as a (one-bit) \textbf{private estimation scheme} (an $ (\epsilon,\delta)$-LDP scheme or a $\gamma$-ML scheme).
The quality of a private estimation scheme is measured by the \textbf{estimation error} which is the worst-case mean squared error (MSE),
\begin{equation}
    R_{n,v}(Q_1,\ldots,Q_n,\hat{\theta}_n) := \sup\limits_{\theta \in \mathcal{P}([v])}\mathbb{E}\left[ \norm{\theta-\hat{\theta}_n(Y^n)}_2^2 \right].
\end{equation}

In this setup, there inherently exists a trade-off between the amount of leakage of private information and the estimation error.
We call this the \textbf{privacy-utility trade-off} (PUT) (under the one-bit communication constraint).
The PUT in our model is defined as the smallest worst-case MSE.
These are defined precisely as follows:
\begin{align}
   \!\!\! \mathrm{PUT}^{\mathrm{LDP}}_n(v,\epsilon,\delta) &\!:= \!\inf\limits_{(Q_1,\ldots,Q_n,\hat{\theta}_n)} R_{n,v}(Q_1,\ldots,Q_n,\hat{\theta}_n),
   \\\!\!\! \mathrm{PUT}^{\mathrm{ML}}_n(v,\gamma) &\!:=\!  \inf\limits_{(Q_1,\ldots,Q_n,\hat{\theta}_n)}R_{n,v}(Q_1,\ldots,Q_n,\hat{\theta}_n),
\end{align}
where the infima are taken over all $(\epsilon,\delta)$-LDP schemes and $\gamma$-ML schemes, respectively.
For simplicity, we will write $\mathrm{PUT}_n$ as one of $\mathrm{PUT}^{\mathrm{LDP}}_n$ or $\mathrm{PUT}^{\mathrm{ML}}_n$ for a statement that does not depend on the choice of the privacy constraint.
We will also often omit the arguments $v,\epsilon,\delta$ and $\gamma$ from $\mathrm{PUT}_n$. 
We will show in what follows that $\mathrm{PUT}_n$ is of the order $\Theta(1/n)$.
Thus, we consider the so-called {\em first-order asymptotics}, i.e.,
\begin{equation}
    \mathrm{PUT} := \liminf\limits_{n \rightarrow \infty} n \cdot \mathrm{PUT}_n.
\end{equation}
 $\mathrm{PUT}$ is a function of the alphabet size $v$ and the parameters that define the privacy constraint, either $(\epsilon,\delta)$ or $\gamma$.
A sequence of private estimation schemes $\{(Q_1,\ldots,Q_n,\hat{\theta}_n)\}_{n=1}^\infty$ is \emph{(asymptotically) optimal} or \emph{achieves $\mathrm{PUT}$} if 
\begin{equation}
    \limsup\limits_{n\rightarrow \infty} n\cdot R_{n,v}(Q_1,\ldots,Q_n,\hat{\theta}_n) = \mathrm{PUT}.
\end{equation}
\begin{figure*}[!ht]
    \begin{equation} \label{eq:PUT_LDP}
        \mathrm{PUT}^{\mathrm{LDP}}(v,\epsilon,\delta)
        \\ = \begin{cases}
        \frac{(v-1)^2}{v} \left( \frac{e^\epsilon+1}{e^\epsilon+2\delta-1} \right)^2 & \text{if }v=\text{even}, \epsilon \geq \zeta (v,\delta)
        \\  \frac{(v-1)^2}{v} \cdot  \frac{(e^\epsilon+1)^2+ \frac{4}{v^2-1} (e^\epsilon+\delta)(1-\delta)}{(e^\epsilon+2\delta-1)^2}  & \text{if } v=\text{odd}, \epsilon \geq \zeta (v,\delta) 
        \\ \frac{(v-1)(v-\delta)}{v \delta} &\text{otherwise}
        \end{cases}.
    \end{equation}
    \hrulefill
\end{figure*}

\section{Main Result}\label{sec:result}
The main contributions of our work are closed-form characterizations of $\mathrm{PUT}^{\mathrm{LDP}}$ and $\mathrm{PUT}^{\mathrm{ML}}$, and the designs and analyses of optimal schemes that achieve the $\mathrm{PUT}$s.

\begin{theorem}\label{thm:main}
    For any $v \geq 2$, $\epsilon>0$, $\delta \in [0,1]$,  and $\gamma  \in (0,\log 2]$,  $\mathrm{PUT}^{\mathrm{LDP}}(v,\epsilon,\delta)$ is characterized as in~\eqref{eq:PUT_LDP}, and
    \begin{equation} \label{PUTML}
        \mathrm{PUT}^{\mathrm{ML}}(v,\gamma) = 
            \frac{(v-1)(v-e^\gamma+1)}{v(e^\gamma-1)},
    \end{equation}
    where
    \begin{align}
        \zeta (v,\delta) &= \log \left( 1 + \frac{2\left(\sqrt{\delta(v^*-1)(v^*-\delta)}-\delta \right)}{v^*} \right),\label{eq:zeta}
    \end{align}
    and
    \begin{equation}
        v^* = 2\left\lceil \frac{v}{2} \right\rceil.
    \end{equation}
\end{theorem}

\begin{figure}[!ht]
     \centering
     \begin{subfigure}[b]{0.4\textwidth}
         \centering
         \includegraphics[width=\textwidth]{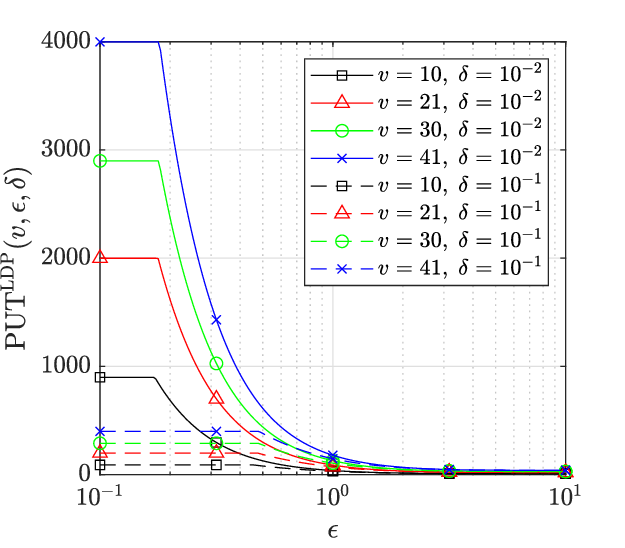}
         \caption{Plot of $\mathrm{PUT}^{\mathrm{LDP}}$}
     \end{subfigure}
     \begin{subfigure}[b]{0.4\textwidth}
         \centering
         \includegraphics[width=\textwidth]{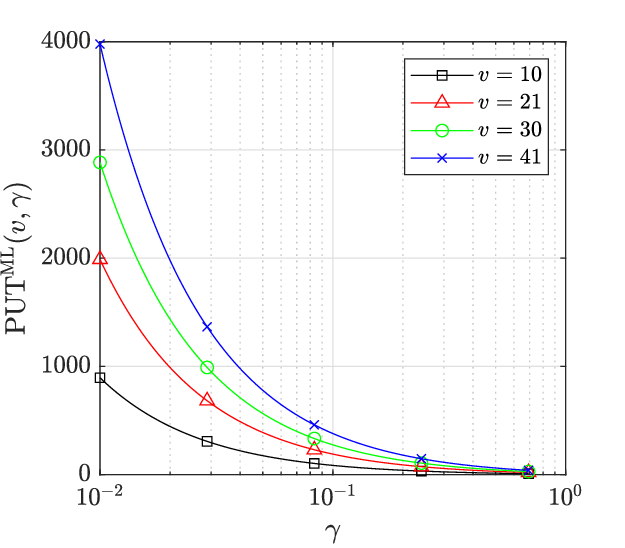}
         \caption{Plot of $\mathrm{PUT}^{\mathrm{ML}}$}
     \end{subfigure}
     \caption{Plots of $\mathrm{PUT}^{\mathrm{LDP}}$ and $\mathrm{PUT}^{\mathrm{ML}}$ in Theorem~\ref{thm:main}. The corners of the lines in (a) correspond to $\epsilon = \zeta(v,\delta)$. The lines in (b) end at $\gamma = \log 2$, where the $\gamma$-ML constraint becomes vacuous.}
     \label{fig:PUT}
\end{figure}
\begin{figure}[!ht]
    \centering
    \includegraphics[width=.4\textwidth]{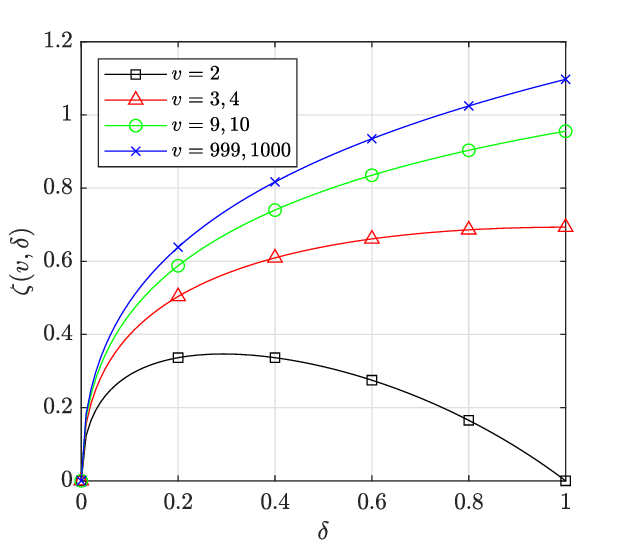}
    \caption{Plot of $\zeta$ in Theorem~\ref{thm:main}.}
    \label{fig:zeta}
\end{figure}

The $\mathrm{PUT}$s and $\zeta$ are depicted in Fig.~\ref{fig:PUT} and~\ref{fig:zeta}, respectively.
As one can naturally expect, the 
$\mathrm{PUT}$s increase in the size of the alphabet of the discrete distribution $v$, and decrease in the parameters for privacy constraints $\epsilon$, $\delta$, and $\gamma$.
Also, $\mathrm{PUT}^{\mathrm{LDP}}$ remains constant for $\epsilon < \zeta(v,\delta)$, i.e., the last case of~\eqref{eq:PUT_LDP}. Note that $\zeta(v, 0)=0$ and thus this case does not occur when $\delta=0$, i.e., pure $\epsilon$-LDP constraint.
This threshold value $\zeta(v,\delta)$ increases in $v$ for all $\delta \in [0,1]$, and increases in $\delta$ for $v \geq 3$.
For $v=2$, $\zeta(2,\delta)$ increases in $\delta$ for $\delta \leq 1-1/\sqrt{2}$ and decreases for $\delta > 1-1/\sqrt{2}$.
On the other hand, note that when $\delta = 1$ or $\gamma = \log 2$, both the $(\epsilon,\delta)$-LDP and the $\gamma$-ML constraints become vacuous.
Thus, the first-order asymptotics of the minimax estimation error (with respect to MSE) under the one-bit communication constraint directly follows from our result as a special case, which is equal to $(v-1)^2 /v$. % the result  $(v-1)^2 /v$

In the rest of the paper, we will prove Theorem~\ref{thm:main} as follows.
For the converse parts, we show that $\mathrm{PUT}$ is asymptotically lower bounded by the PUT of the another model $\mathrm{PUT}_{\mathrm{SR}}$ which exploits i.i.d.\ {\em shared randomness} between the clients and the server.
Next, we derive a lower bound on $\mathrm{PUT}_{\mathrm{SR}}$ by exploiting local asymptotic normality \cite{ibragimov13_stat_asymp, le00_asymptotics, van98_asymptotic} based on the results by Ye and Barg~\cite{ye17_opt_PUT_l2}.
The lower bound can be tightened by maximizing a convex function defined on $\mathcal{Q}$.
By characterizing the set of all extreme points of $\mathcal{Q}$ and solving the resultant optimization problem, we obtain the desired lower bounds.
For the achievability parts, we first construct optimal schemes for the model with shared randomness achieving $\mathrm{PUT}_{\mathrm{SR}}$, whose privacy mechanisms are appropriate modifications of the \emph{resolutions of block design (or RPBD) mechanisms} proposed in \cite{park23_block,nam23_res_BD}, for some cases.
The corresponding estimators are also judiciously designed and are  distinguished from the estimators proposed in previous works~\cite{park23_block,nam23_res_BD}.
Finally, we construct optimal schemes for our model so that in the limit of a large number of clients $n$, they resemble the optimal schemes for the model with shared randomness.

\section{Model with Shared Randomness}\label{sec:model_SR}

\begin{figure}[!ht]
    \centering
    \includegraphics[width=0.6\linewidth]{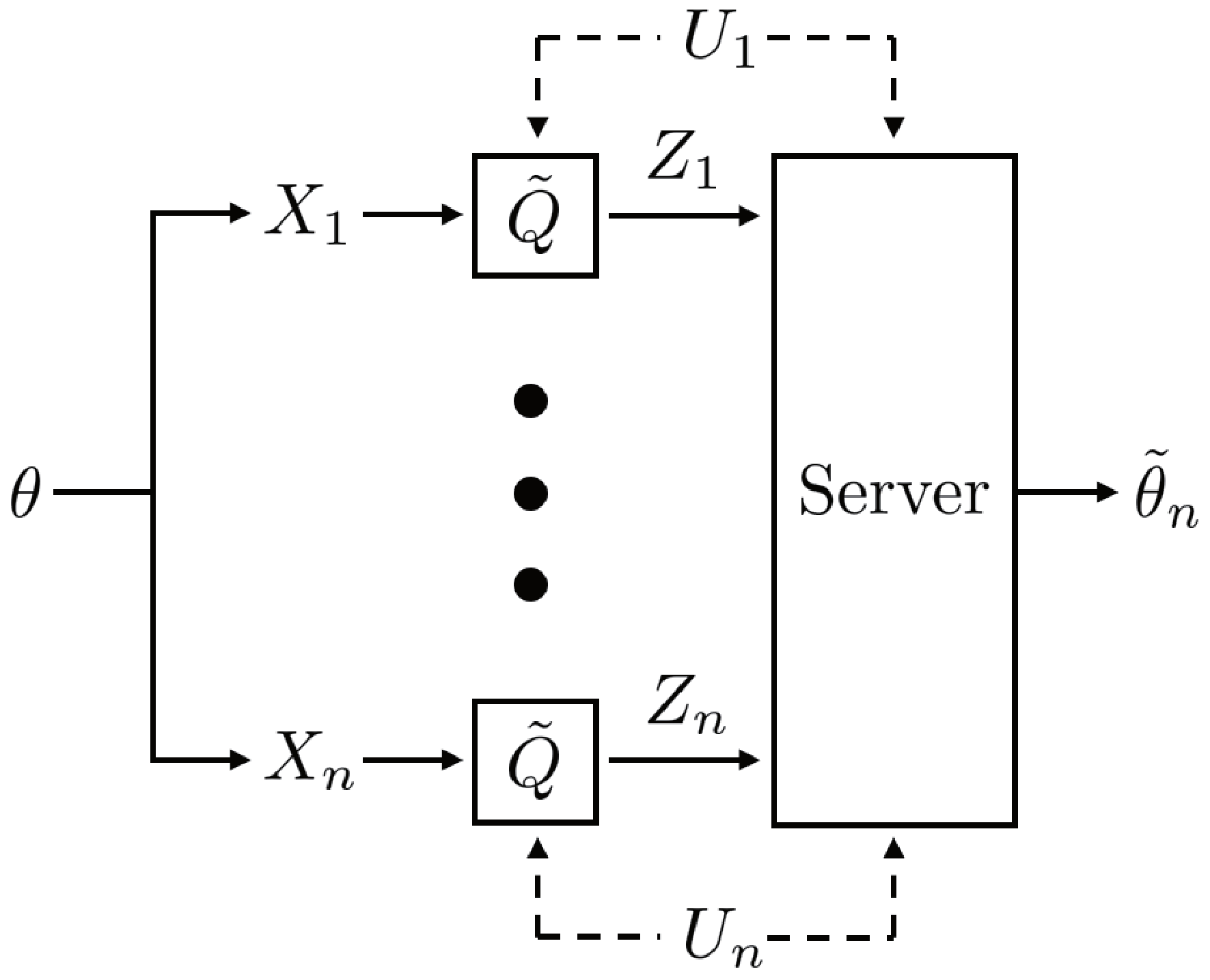}
    \caption{Model with shared randomness}
    \label{fig:model_SR}
\end{figure}

We prove Theorem~\ref{thm:main} by demonstrating an equivalence between the $\mathrm{PUT}$ of our model and the $\mathrm{PUT}_{\mathrm{SR}}$ of another model with (i.i.d.) shared randomness.
In this section, we define the model with shared randomness precisely.
The setup is depicted in Fig.~\ref{fig:model_SR}.
The main difference to the original model is that for all $i \in [n]$, the server and the $i$-th client have access to a shared randomness $U_i \in \mathcal{U}$, $|\mathcal{U}|<\infty$, in advance.
We assume that $U_1,\ldots,U_n$ are i.i.d.\ random variables with $U_i \sim P_U \in \mathcal{P}(\mathcal{U})$, and all the clients and the server can pre-determine $P_U$ for generating $U^n$, in advance.
Also, we assume that $U^n$ and $X^n$ are independent.
Then, each of the $n$ clients perturbs its data $X_i$ through a conditional probability mass function $\tilde{Q}:\mathcal{U} \times \mathcal{X} \rightarrow \mathcal{Z}$ where $|\mathcal{Z}|<\infty$, with the knowledge of the shared randomness $U_i$, i.e., for given $U_i=u_i$ and $X_i=x_i$, $Z_i$ is sampled from $\tilde{Q}(\cdot|u_i,x_i)$.
For all $u \in \mathcal{U}$, we denote $\mathcal{Z}_{u}$ as
\begin{equation}
    \mathcal{Z}_{u} := \{z \in \mathcal{Z}: \tilde{Q}(z|u,x) > 0 \text{ for some }x\}.
\end{equation}
In this model, the constraints are slightly modified so that $\tilde{Q}$ should satisfy the constraints for any given realization of shared randomness $U$.
\begin{definition}\label{def:priv_SR}
    For $\epsilon > 0$ and $\delta \in [0,1]$, a pair $(P_U,\tilde{Q})$ is called a (one-bit) $(\epsilon,\delta)$-LDP mechanism with shared randomness if
    \begin{equation}
        \forall u \in \mathcal{U}, \quad \tilde{Q}(\cdot|u,\cdot) \in \mathcal{Q}^{(\epsilon,\delta)}.
    \end{equation}
    For $\gamma \in (0,\log 2]$, a pair $(P_U,\tilde{Q})$ is called a (one-bit) $\gamma$-ML mechanism with shared randomness if
    \begin{equation}
        \forall u \in \mathcal{U}, \quad \tilde{Q}(\cdot|u,\cdot) \in \mathcal{Q}^{\gamma}.
    \end{equation}
\end{definition}
For notational simplicity, we define $\tilde{\mathcal{Q}}^{(\epsilon,\delta)}$ as the set of all $(\epsilon,\delta)$-LDP mechanisms with shared randomness and $\tilde{\mathcal{Q}}^{\gamma}$ as the set of all $\gamma$-ML mechanisms with shared randomness.

After perturbing the data $X^n$ into $Z^n$, the server collects $Z^n$ and estimates $\theta$ with the knowledge of the shared randomness $U^n$ using the estimator $\tilde{\theta}_n:\mathcal{U}^n \times \mathcal{Z}^n \rightarrow \mathbb{R}^v$.
We denote a tuple of a privacy mechanism with shared randomness and an estimator, $(P_U,\tilde{Q},\tilde{\theta}_n)$ as a \emph{private estimation scheme with shared randomness} (an $(\epsilon,\delta)$-LDP scheme with shared randomness or a $\gamma$-ML scheme with shared randomness).
The estimation error of a private estimation scheme with shared randomness is also defined to be the worst-case MSE,
\begin{equation}
    R_{n,v}(P_U,\tilde{Q},\tilde{\theta}_n):=\sup\limits_{\theta \in \mathcal{P}([v])}\mathbb{E}\left[\norm{\theta-\tilde{\theta}_n(U^n,Z^n)}_2^2\right].
\end{equation}
The PUTs in this model are defined as
\begin{align}
    \mathrm{PUT}_{\mathrm{SR},n}^{\mathrm{LDP}}(v,\epsilon,\delta) &:= \inf\limits_{(P_U,\tilde{Q},\tilde{\theta}_n)}R_{n,v}(P_U,\tilde{Q},\tilde{\theta}_n),
    \\ \mathrm{PUT}_{\mathrm{SR},n}^{\mathrm{ML}}(v,\gamma) &:= \inf\limits_{(P_U,\tilde{Q},\tilde{\theta}_n)}R_{n,v}(P_U,\tilde{Q},\tilde{\theta}_n),
\end{align}
where the infima are taken over all $(\epsilon,\delta)$-LDP schemes with shared randomness and $\gamma$-LDP schemes with shared randomness, respectively.
For simplicity, we omit the upper indices of $\mathrm{PUT}_{\mathrm{SR},n}$ and $\tilde{\mathcal{Q}}$ with the same convention as $\mathrm{PUT}_n$ and $\mathcal{Q}$.
We will also often omit the arguments $v,\epsilon,\delta$, and $\gamma$ from $\mathrm{PUT}_{\mathrm{SR},n}$. 
The  {\em first-order asymptotics} of $\mathrm{PUT}_{\mathrm{SR},n}$ is defined as
\begin{equation}
    \mathrm{PUT}_{\mathrm{SR}} := \liminf_{n\rightarrow \infty} n \cdot \mathrm{PUT}_{\mathrm{SR},n}.
\end{equation}
We say that a sequence of private estimation schemes with shared randomness $\{(P_U,\tilde{Q},\tilde{\theta}_n)\}_{n=1}^\infty$ is \emph{(asymptotically) optimal} or \emph{achieves $\mathrm{PUT}_{\mathrm{SR}}$} if 
\begin{equation}
    \limsup\limits_{n\rightarrow \infty}  n \cdot R_{n,v}(P_U,\tilde{Q},\tilde{\theta}_n) = \mathrm{PUT}_{\mathrm{SR}}.
\end{equation}

\section{Converse}\label{sec:conv}
In this section, we prove the converse part of Theorem~\ref{thm:main}.
At first, we prove $\mathrm{PUT} \geq \mathrm{PUT}_{\mathrm{SR}}$.
Then, we derive a lower bound of $\mathrm{PUT}_{\mathrm{SR}}$ by exploiting local asymptotic normality \cite{ye17_opt_PUT_l2, le00_asymptotics,ibragimov13_stat_asymp,van98_asymptotic}.
Because the derived lower bound is related to the maximum of a convex function defined on $\mathcal{Q}$, we obtain the tightest lower bound by characterizing all the extreme points of $\mathcal{Q}$, which is a bounded convex set.

\subsection{Comparing models: Converse}
We show that $\mathrm{PUT}$ is lower bounded by $\mathrm{PUT}_{\mathrm{SR}}$.
\begin{proposition}\label{prop:model_equiv_conv}
It holds that
    \begin{equation}
        \mathrm{PUT} \geq \mathrm{PUT}_{\mathrm{SR}}.
    \end{equation}
\end{proposition}
\begin{IEEEproof}
    For any given $n \in \mathbb{N}$ and a private estimation scheme $(Q_1,\ldots,Q_n, \hat{\theta}_n)$, we construct a sequence of private estimation schemes with shared randomness $\{(P_U,\tilde{Q},\tilde{\theta}_m)\}_{m=1}^\infty$ as follows:
    First, we construct $(P_U,\tilde{Q})$ as
    \begin{equation}
        \mathcal{U} = [n], \quad P_U = \mathrm{Unif}(\mathcal{U}),\quad \mathcal{Z} = \bigcup\limits_{i=1}^n \mathcal{Y}_i,
    \end{equation}
    \begin{equation}
        \forall u \in \mathcal{U}, z\in \mathcal{Y}_{u}, x \in \mathcal{X},\quad \tilde{Q}(z|u,x) = Q_u(z|x).
    \end{equation}
    Clearly, $(P_U,\tilde{Q}) \in \tilde{\mathcal{Q}}$.
    Now, let $T:\mathcal{U}^m \rightarrow \mathbb{Z}_{\geq 0}$,
    \begin{equation}
        T(u^m) = \min\limits_{ j \in \mathcal{U}}  \sum_{i=1}^{m} \mathbbm{1}(u_i = j),
    \end{equation}
    which denotes the minimum number of occurrences of a symbol in the vector $u^m = (u_1, \ldots, u_m)$.
    % Then, for any given $U^m = u^m$, $m$ clients can be rearranged so that each of the $(i-1)n+1, \ldots,in$-th clients perturbs its data through $\tilde{Q}(\cdot|1,\cdot),\ldots,\tilde{Q}(\cdot|n,\cdot)$, for all $i \in [T(u^m)]$. 
    Then, for any $u^m \in \mathcal{U}^m$, there are $T(u^m)$ vectors $\tau_1,\ldots,\tau_{T(u^m)} \in [m]^n$ such that $u_{\tau_i} = (1,\ldots,n)$ and all elements of $\tau_i$ are distinct for every $i \in [T(u^m)]$, and $\tau_i,\tau_j$ have no common element for all $i\neq j$.
    We fix a deterministic rule that assigns such $\tau_1,\ldots,\tau_{T(u^m)}$ for each $u^m \in \mathcal{U}^m$ satisfying $T(u^m) \geq 1$.
    Next, we define the estimator $\tilde{\theta}_m$ as
    \begin{equation}
        \!\!\tilde{\theta}_m(u^m,z^m)
         \!=\! \begin{cases}
            \mathbf{0} & \text{if }T(u^m)\!=\! 0
            \\ \frac{1}{T(u^m)} \sum\limits_{i=1}^{T(u^m)} \hat{\theta}_n(z_{\tau_i}) & \text{otherwise}
        \end{cases}.
    \end{equation}

    Up to this point, we constructed a private estimation scheme with shared randomness $(P_U,\tilde{Q},\tilde{\theta}_m)$ based on a given private estimation scheme $(Q_1,\ldots,Q_n,\hat{\theta}_n)$.
    Next, we compare their estimation errors.
    % \begin{lemma}[Hoeffding's inequality \cite{Hoeffding}]\label{lem:Hoeff}
    % If $X_1,\ldots,X_n$ are independent real random variables with $\mathbb{E}[X_i] = \mu_i$ and $X_i \in [a,b]$ for all $i\in[n]$, then for $t > 0$,
    % \begin{equation}
    %     \mathrm{Pr} \left(\frac{1}{n}\sum\limits_{i=1}^n (X_i - \mu_i) \geq t \right) \leq \exp\left(-\frac{2 t^2}{(b-a)^2} n \right).
    % \end{equation}
    % \end{lemma}
    Let $\delta=n^{-2}$.
    Because $P_U = \mathrm{Unif}([n])$, the union bound and Hoeffding's inequality~\cite{Hoeffding} yield
    \begin{align}\label{eq:Hoeff1}
        \mathrm{Pr}&\left( T(U^m) \leq m\left( \frac{1}{n} - \delta \right) \right)\nonumber
        \\& \leq \sum\limits_{j=1}^n \mathrm{Pr}\left( \sum\limits_{i=1}^m \mathbbm{1}(U_i = j) \leq m\left( \frac{1}{n} - \delta \right) \right)
        \\& \leq n \exp (-2 m/n^4). \label{eq:Hoeff2}
    \end{align}
    For $(u^m,z^m)$ such that $T(u^m) \geq 1$, we denote
    \begin{equation}
        \tilde{L}(u^m,z^m) = \norm{\frac{1}{T(u^m)} \sum\limits_{i=1}^{T(u^m)} (\theta - \hat{\theta}_n(z_{\tau_i}))}_2^2.
    \end{equation}
    Then,~\eqref{eq:Hoeff2} implies that 
    \begin{align} \label{eq:Conv_typ_1}
        &\mathbb{E}\left[\norm{\theta - \tilde{\theta}_m(U^m,Z^m)}_2^2\right] \nonumber
        \\& \leq \mathbb{E}\left[ \tilde{L}(U^m,Z^m) \middle| T(U^m) > \frac{m(n-1)}{n^2} \right] 
        \\& +  n \exp(-2 m / n^4)\times \nonumber
        \\&\quad\left( \mathbb{E}\left[ \tilde{L}(U^m,Z^m) \middle| T(U^m) \in \left[1,\frac{m(n-1)}{n^2}\right] \right] + 1 \right), \nonumber
    \end{align}
    because $\|\theta\|_2^2 \leq 1$.
    Next, let $L(\theta) = \mathbb{E}\big[\|\theta-\hat{\theta}_n(Y^n)\|_2^2\big]$.
    Note that for any given $U^m=u^m$ satisfying $T(u^m) \geq 1$,  $Z_{\tau_i}$ and $Y^n$ follow the same distribution by the construction, and $Z_{\tau_1},\ldots,Z_{\tau_{T(u^m)}}$ are mutually independent.
    Thus, for $t \in \left[ \lfloor m/n \rfloor \right]$, we have
    \begin{multline}
        \mathbb{E}\left[ \tilde{L}(U^m,Z^m) \middle| T(U^m) =t \right]
        \\ =  \frac{1}{t^2} \sum\limits_{i=1}^t \mathbb{E}\left[\norm{ (\theta - \hat{\theta}_n(Z_{\tau_i}))}_2^2  \middle| T(U^m) = t \right]  = \frac{L(\theta)}{t}.
    \end{multline}
    Using this fact,~\eqref{eq:Conv_typ_1} yields
    \begin{multline}
        \mathbb{E}\left[\norm{\theta - \tilde{\theta}_m(U^m,Z^m)}_2^2\right] 
        \\ \leq \frac{n^2 L(\theta)}{m(n-1)} +  n \exp(-2 m / n^4)\left(L(\theta) + 1 \right).
    \end{multline}
    By taking the supremum over $\theta \in \mathcal{P}([v])$ on both sides and using the fact that $(P_U,\tilde{Q},\tilde{\theta}_m)$ is just a special case of a private estimation scheme with shared randomness, we obtain
    \begin{multline}
        \mathrm{PUT}_{\mathrm{SR},m} \leq \frac{n^2}{m(n-1)}\sup\limits_{\theta \in \mathcal{P}([v])}L(\theta)
        \\ +  n \exp(-2 m / n^4)\left(\sup\limits_{\theta \in \mathcal{P}([v])}L(\theta) + 1 \right).
    \end{multline}
    Next, by multliplying $m$ and taking $\liminf_{m\rightarrow \infty}$ on  both sides, we have
    \begin{equation}
        \mathrm{PUT}_{\mathrm{SR}}  \leq \frac{n^2}{n-1}\sup\limits_{\theta \in \mathcal{P}([v])}L(\theta).
    \end{equation}
    Because the above inequality holds for any $n \in \mathbb{N}$ and any private estimation scheme $(Q_1,\ldots,Q_n,\hat{\theta}_n)$, we can get the desired result.
\end{IEEEproof}

\subsection{Local asymptotic normality} \label{sec:LAN}
In this subsection, we derive a lower bound on $\mathrm{PUT}_{\mathrm{SR}}$ by exploiting the local asymptotic normality property as was done in~\cite{ye17_opt_PUT_l2}.
For the model with shared randomness in Section~\ref{sec:model_SR}, the server receives i.i.d.\ random variables $W_1,\ldots,W_n$ of the form $W_i = (U_i,Z_i)$, each following the distribution $P_W^\theta(u,z) = \sum\limits_{x\in \mathcal{X}} Q(u,z|x)\theta_x$ where $Q(u,z|x) := P_U(u)\tilde{Q}(z|u,x)$.
Here, $Q:\mathcal{X} \rightarrow\mathcal{P}(\mathcal{W})$ satisfies the privacy constraint ($(\epsilon,\delta)$-LDP or $\gamma$-ML) because $(P_U,\tilde{Q}) \in \tilde{\mathcal{Q}}$, but it is only guaranteed that $|\mathcal{W}|<\infty$ instead of satisfying the one-bit communication constraint.
Accordingly, we denote $\mathcal{Q}_*$ as the set of all privacy mechanisms ($(\epsilon,\delta)$-LDP mechanisms or $\gamma$-ML mechanisms) $Q:\mathcal{X} \rightarrow \mathcal{P}(\mathcal{W})$ such that $Q = P_U \tilde{Q}$ for some $(P_U,\tilde{Q})\in\tilde{\mathcal{Q}}$.
Note that $\mathcal{P}([v])$ is a $(v-1)$-dimensional manifold.
Hence, we choose a coordinate function $\varphi:\mathcal{P}([v]) \rightarrow \mathbb{R}^{v-1}$, $\varphi_i(\theta) = \theta_i$.
Let $\Phi \in \mathbb{R}^{v-1}$ be a random variable following a uniform prior distribution $\lambda$ supported on the small neighborhood of $\mathbf{1}_{(v-1)} / v$.
More precisely, $\lambda$ is supported on the $(v-1)$-dimensional ellipsoid $
\big\{ \varphi \!:\! \sum_{i=1}^{v-1} (\varphi_i - \frac{1}{v} )^2  +( \sum_{i=1}^{v-1}\ (\varphi_i - \frac{1}{v}))^2 < n^{-10/13}  \big\}$.
Then, we have
\begin{align}
    \sup\limits_{\theta \in \mathcal{P}([v])} \mathbb{E} & \left[ \norm{\theta  - \tilde{\theta}_n(W^n)}_2^2 \right] \nonumber
     \\&\geq \mathbb{E}_{\Phi \sim \lambda}\left[ \norm{J^\top \Phi + e_v -\tilde{\theta}_n(W^n)}_2^2 \right],
\end{align}
where $J = (I_{(v-1)\times (v-1)} , -\mathbf{1}_{(v-1)})$, and $e_v$ is the $v$-dimensional vector $e_v = (0,\ldots,0,1)$.
Because the posterior mean of $J^\top \Phi + e_v$, which is also the Bayes estimator, minimizes the MSE,
we have
\begin{multline}
    \inf\limits_{\tilde{\theta}_n}\sup\limits_{\theta \in \mathcal{P}([v])} \mathbb{E}\left[ \norm{\theta - \tilde{\theta}_n(W^n)}_2^2 \right]
    \\ \geq \mathbb{E}\left[ \mathrm{Tr}(\mathrm{Cov}(\Phi|W^n)) + \mathbf{1}^\top \mathrm{Tr}(\mathrm{Cov}(\Phi|W^n))\mathbf{1} \right].
\end{multline}
The local asymptotic normality property~\cite{ibragimov13_stat_asymp, le00_asymptotics, van98_asymptotic} implies that the posterior distribution of $\Phi$ given $W^n$ converges to the Normal distribution with mean $\mathbf{1}/v$ and covariance $\frac{1}{n}J^\top I_W^{-1}(\mathbf{1}/v) J$, where $I_W$ denotes the Fisher information matrix,
\begin{align}
    \!\!  I_W(\varphi)\! =\!  \mathbb{E}\left[ \left(\!  \pdv{}{\varphi} \log P_W^\theta(W) \right)\left( \! \pdv{}{\varphi} \log P_W^\theta (W) \right)^\top \right].\! 
\end{align}
With this idea, \cite[Sec.~V]{ye17_opt_PUT_l2} derived a lower bound which holds uniformly for all $Q \in \mathcal{Q}_*$: There exist positive constants $C_1,C_2$ and an integer $N$ such that for all $Q \in \mathcal{Q}_*$,
\begin{multline}\label{eq:Ye_LAN}
    \inf\limits_{\tilde{\theta}_n}\sup\limits_{\theta \in \mathcal{P}([v])} \mathbb{E}\left[ \norm{\theta - \tilde{\theta}_n(W^n)}_2^2 \right]  \geq \frac{1}{n}\left( 1 - \frac{C_1}{n^{1/13}} \right)\times
    \\ \sup\limits_{Q \in \mathcal{Q}_*}\left( \mathrm{Tr}(I_W^{-1}(\mathbf{1}/v)) + \mathbf{1}^\top I_W^{-1}(\mathbf{1}/v) \mathbf{1} \right)- \frac{C_2}{n^{14/13}},
\end{multline}
whenever $n \geq N$.\footnote{In \cite{ye17_opt_PUT_l2}, the authors only considered the $\epsilon$-LDP constraint.
However,~\eqref{eq:Ye_LAN} also holds uniformly for all $Q \in \mathcal{Q}_*$ because its proof does not rely on the choice of privacy constraint, apart from the inequalities between (72) and (73) in~\cite{ye17_opt_PUT_l2}.
To check the validity of~\eqref{eq:Ye_LAN}, it is sufficient to check that $|Q(w|x)-Q(w|x')|/\sum_{x\in\mathcal{X}}Q(w|x)$ is bounded for all $x,x'\in\mathcal{X},w \in \mathcal{W}$; this is, however, easy to verify.
}
Thus, we obtain
\begin{equation}
    \mathrm{PUT}_{\mathrm{SR}} \geq \sup\limits_{Q \in \mathcal{Q}_*}\left( \mathrm{Tr}(I_W^{-1}(\mathbf{1}/v)) + \mathbf{1}^\top I_W^{-1}(\mathbf{1}/v) \mathbf{1} \right).
\end{equation}
By applying \cite[Prop.~V.12]{ye17_opt_PUT_l2} and some further manipulations \cite[Eq.~(79)]{ye17_opt_PUT_l2}, we have
\begin{equation}
    \mathrm{PUT}_{\mathrm{SR}} \geq \frac{(v-1)^2}{v\Big(\sup\limits_{Q \in \mathcal{Q}_*}F(Q) - 1\Big)}, \label{eq:Ye_lem}
\end{equation}
where
\begin{equation}
    F(Q) = \sum\limits_{w \in \mathcal{W}} \frac{\sum\limits_{x\in \mathcal {X}} \left(Q(w|x)\right)^2 }{\sum\limits_{x\in \mathcal {X}} Q(w|x)}.
\end{equation}

We modify the right-hand side (RHS) of~\eqref{eq:Ye_lem} to get a bound that is related to $\mathcal{Q}$ instead of $\mathcal{Q}_*$.
\begin{lemma}\label{lem:LAN_LB}
It holds that
    \begin{equation}
        \mathrm{PUT}_{\mathrm{SR}} \geq \frac{(v-1)^2}{v \left( \sup\limits_{Q\in\mathcal{Q}} F(Q)-1 \right)}.
    \end{equation}
\end{lemma}
\begin{IEEEproof}
    For any given $(\tilde{Q},P_U) \in \tilde{\mathcal{Q}}$ and $Q = P_U \tilde{Q}$,
    \begin{multline}
        F(Q) = \sum\limits_{u\in \mathcal {U}} P_U(u) \sum\limits_{z \in \mathcal{Z}} \frac{\sum\limits_{x\in \mathcal {X}} (\tilde{Q}(z|u,x))^2}{\sum\limits_{x\in \mathcal {X}} \tilde{Q}(z|u,x)}
        \\ \leq \sup\limits_{u\in \mathcal {U}} \sum\limits_{z \in \mathcal{Z}} \frac{\sum\limits_{x\in \mathcal {X}} (\tilde{Q}(z|u,x))^2}{\sum\limits_{x\in \mathcal {X}} \tilde{Q}(z|u,x)} \leq \sup\limits_{Q \in \mathcal{Q}} F(Q),
    \end{multline}
    because $(P_U,\tilde{Q}) \in \tilde{\mathcal{Q}}$ implies $\tilde{Q}(\cdot|u,\cdot) \in \mathcal{Q}$ for all $u \in \mathcal{U}$ by Definition~\ref{def:priv_SR}.
    By plugging above inequalities into~\eqref{eq:Ye_lem}, we get the desired result.
\end{IEEEproof}

\subsection{Extreme points of privacy mechanisms}\label{sec:ex_pt}
In the previous subsection, we derive a lower bound on  $\mathrm{PUT}_{\mathrm{SR}}$ as in Lemma~\ref{lem:LAN_LB}.
To obtain closed-form lower bounds, it remains to solve the optimization problem $\sup_{Q \in \mathcal{Q}} F(Q)$.
Note that $F(Q) = F(Q')$ if $Q \cong Q'$.
Thus, in the remaining part of this section, we treat $Q \in \mathcal{Q}$ as a (row) stochastic matrix in $[0,1]^{v \times 2}$ without loss of generality.
It can be easily checked that $F$ is a convex function on $\mathcal{Q}$, and $\mathcal{Q}$ is a bounded convex set (cf. \cite{kairouz14_extremal}, \cite{holohan17_LDP_ex_pts}).
Thus, the supremum is achieved at an extreme point of $\mathcal{Q}$.
Accordingly, we characterize all the extreme points of $\mathcal{Q}$ and solve $\sup_{Q \in \mathcal{Q}} F(Q)$ by comparing the values of $F(Q)$ at the extreme points.
\begin{proposition}\label{prop:ex_pt}
    A stochastic matrix $Q \in [0,1]^{v\times 2}$ is an extreme point of $\mathcal{Q}^{(\epsilon,\delta)}$ if and only if $Q$ has a column contained in
    \begin{equation}\label{eq:LDP_ex_col}
        \left\{\frac{e^\epsilon+\delta}{e^\epsilon+1},\frac{1-\delta}{e^\epsilon+1}\right\}^v \cup \{\delta,0\}^v \cup \{\mathbf{0}\}.
    \end{equation}
    In addition, a stochastic matrix $Q \in [0,1]^{v\times 2}$ is an extreme point of $\mathcal{Q}^{\gamma}$ if and only if $Q$ has a column contained in $\{{e^\gamma-1}, 0\}^v \cup \{\mathbf{0}\} $.
\end{proposition}

\begin{IEEEproof}
    Note that $Q \in \mathcal{Q}$ is a convex combination of $Q',Q'' \in \mathcal{Q}$ if and only if the first column of $Q$ is a convex combination of the first columns of $Q',Q''$, because the second column is just $\mathbf{1}$ minus the first column.
    Thus, we focus on the first column of the privacy mechanisms.

    We first focus on the LDP constraint.
    Let $q$ be the first column of $Q \in \mathcal{Q}^{(\epsilon,\delta)}$, and $m = \min_x q_x$, $M=\max_x q_x$.
    By definition, $Q \in \mathcal{Q}^{(\epsilon,\delta)}$ if and only if
    \begin{align}\label{eq:LDP_ex_cond}
        &e^\epsilon m + \delta \geq M, \quad e^\epsilon(1-M) + \delta \geq 1-m, \nonumber
        \\& 0 \leq m \leq M \leq 1.
    \end{align}
    Let $\mathcal{M}$ denote the set of $(m,M)$ satisfying~\eqref{eq:LDP_ex_cond}, which is a bounded convex polytope.
    The extreme points of $\mathcal{M}$ are
    \begin{equation}\label{eq:LDP_ex_pt}
        (0,0),\; (0,\delta),\;\left( \frac{1-\delta}{e^\epsilon+1},\frac{e^\epsilon+\delta}{e^\epsilon+1}\right),\;  (1-\delta,1),\; (1,1).
    \end{equation}
    Let $Q$ be the stochastic matrix which have a column $q$ in~\eqref{eq:LDP_ex_col}.
    Clearly, $Q \in \mathcal{Q}^{(\epsilon,\delta)}$ because $q$ satisfies~\eqref{eq:LDP_ex_cond}.
    Assume that $q$ is a convex combination of some other vectors $q',q''$.
    Let $x_1 = \argmin_x q_x$, $x_2 = \argmax_x q_x$, and $m',m''$ and $M',M''$ be the $x_1$-th and $x_2$-th entries of $q',q''$, respectively.
    Because $(m,M)$ is a convex combination of $(m',M')$ and $(m'',M'')$, and $(m,M)$ is an extreme point of a bounded convex polytope $\mathcal{M}$, either $(m',M')$ or $(m'',M'')$ is not contained in $\mathcal{M}$.
    Accordingly, either $Q' = (q',\mathbf{1}-q')$ or $Q'' = (q'',\mathbf{1}-q'')$ is not contained in $\mathcal{Q}^{(\epsilon,\delta)}$, and this implies that $Q$ is an extreme point of $\mathcal{Q}^{(\epsilon,\delta)}$.

    It remains to prove the only if part of the proposition.
    We will show that if $Q$ is an extreme point, then $q$ takes at most two values.   
    Suppose that there exists $x^* \in \mathcal{X}$ such that $q_{x^*} \neq m$ and $q_{x^*} \neq M$.
    Then, let $q'$ and $q''$ be the vectors such that
    \begin{equation}
        q'_x = \begin{cases} m & \text{if }x=x^*
            \\ q_x & \text{otherwise} 
        \end{cases}, \quad q''_x = \begin{cases} M & \text{if }x=x^*
            \\ q_x & \text{otherwise}
        \end{cases}.
    \end{equation}
    It can be easily seen that $q$ is the convex combination of $q'$ and $q''$, whose corresponding stochastic matrices $Q' = (q',\mathbf{1}-q')$ and $Q'' = (q'',\mathbf{1}-q'')$ are also in $\mathcal{Q}^{(\epsilon,\delta)}$.
    Thus, $Q$ is an extreme point only if $q \in \{m,M\}^v$.
    Next, we derive a necessary condition on $(m,M)$ when $Q$ is an extreme point.
    If $(m,M)$ is not one of the extreme points $\mathcal{M}$, then it is a convex combination of the points in~\eqref{eq:LDP_ex_pt}.
    Combining with the fact that $q \in \{m,M\}^v$ when $Q$ is an extreme point of $\mathcal{Q}^{(\epsilon,\delta)}$, we can conclude that if $Q$ is an extreme point of $Q^{(\epsilon,\delta)}$, then $Q$ should have a column which is contained in the set in~\eqref{eq:LDP_ex_col}.

    For $\mathcal{Q}^\gamma$, the above proof steps follow {\em mutatis mutandis} apart from the fact that $Q \in \mathcal{Q}^\gamma$ if and only if
    \begin{equation}
        M+1-m \leq e^\gamma, \quad 0 \leq m \leq M \leq 1,
    \end{equation}
    and the extreme points of the set of $(m,M)$ satisfying the above are
    \begin{equation}
         (0,0),\; (0,e^\epsilon-1),\; (2-e^\epsilon, 1 ),\; (1,1).
    \end{equation}
    This completes the proof.
\end{IEEEproof}

Because we characterized all the extreme points of $\mathcal{Q}$, we can get a closed-form expression of $\sup_{Q \in \mathcal{Q}} F(Q)$.
By substituting the optimized values of $F(Q)$ into Lemma~\ref{lem:LAN_LB}, we can get the desired lower bounds of $\mathrm{PUT}_{\mathrm{SR}}$.
\begin{proposition}\label{prop:PUT_SR_LB}
    For any $v \geq 2$, $\epsilon>0$, $\delta \in [0,1]$, and $\gamma  \in (0,\log 2]$, $\mathrm{PUT}_{\mathrm{SR}}^{\mathrm{LDP}}(v,\epsilon,\delta)$ and $\mathrm{PUT}^{\mathrm{ML}}_{\mathrm{SR}}(v,\gamma)$ are lower-bounded by the RHSs of \eqref{eq:PUT_LDP} and \eqref{PUTML}, respectively, where $\zeta(v,\delta)$ is given in~\eqref{eq:zeta}.
%    \begin{equation}\label{eq:PUT_SR_ML}
 %       \mathrm{PUT}^{\mathrm{ML}}_{\mathrm{SR}}(v,\gamma) \geq \frac{(v-1)(v-e^\gamma+1)}{v(e^\gamma-1)},
  %  \end{equation}
   
\end{proposition}
% \begin{figure*}
%     \begin{equation}\label{eq:PUT_SR_LDP}
%         \mathrm{PUT}^{\mathrm{LDP}}_{\mathrm{SR}}(v,\epsilon,\delta) \geq\begin{cases}
%         \frac{(v-1)^2}{v} \left( \frac{e^\epsilon+1}{e^\epsilon+2\delta-1} \right)^2 & \text{if }v=\text{even}, \epsilon \geq \zeta (v,\delta)
%         \\  \frac{(v-1)^2}{v} \cdot  \frac{(e^\epsilon+1)^2+ \frac{4}{v^2-1} (e^\epsilon+\delta)(1-\delta)}{(e^\epsilon+2\delta-1)^2}  & \text{if } v=\text{odd}, \epsilon \geq \zeta (v,\delta) 
%         \\ \frac{(v-1)(v-\delta)}{v \delta} &\text{otherwise}
%         \end{cases}.
%     \end{equation}
%     \hrulefill
% \end{figure*}
\begin{IEEEproof}
    As a first step, we solve $\sup_{Q \in \mathcal{Q}} F(Q)$.
    Because $F$ is a convex function on $\mathcal{Q}$, it is sufficient to optimize $F$ over the extreme points of $\mathcal{Q}$.
    In Proposition~\ref{prop:ex_pt}, we characterized all the extreme points of $\mathcal{Q}$.
    The following lemma simplifies the calculation of $F(Q)$ for such extreme points, and we omit its proof because it can be derived through simple calculations.
    \begin{lemma} \label{lem:stochastic_matrix}
    If a stochastic matrix $Q \in [0,1]^{v\times 2}$ has a column $q \in \{a,1-a\}^v$, $a \in [0,1]$, and $t$ elements of $q$ are $a$, then, 
        \begin{equation}
            F(Q) =  1 + \frac{(2a-1)^2}{a^2 + \frac{t^2 + (v-t)^2}{t(v-t)}a(1-a)   + (1-a)^2}.
        \end{equation}
    If a stochastic matrix $Q \in [0,1]^{v\times 2}$ has a column $q \in \{a,0\}^v$, $a \in [0,1]$, and $t\geq 1$ elements of $q$ are $a$, then,
    \begin{equation}
        F(Q) = 2 - \frac{(1-a)v}{v-at}.
    \end{equation}
    \end{lemma}

    We first focus on the LDP constraint.
    By Lemma~\ref{lem:stochastic_matrix}, we have the following conclusions:
    1) If $Q$ has a zero column, a simple calculation gives $F(Q) = 1$.
    2) Suppose $Q$ has a column $q \in \left\{\frac{e^\epsilon+\delta}{e^\epsilon+1},\frac{1-\delta}{e^\epsilon+1}\right\}^v$ and $t$ elements of $q$ are $\frac{e^\epsilon+\delta}{e^\epsilon+1}$. Then,
    \begin{multline}\label{eq:F_LDP_t}
        F(Q)= 1
        \\  +\frac{(e^\epsilon + 2\delta - 1)^2}{(e^\epsilon+ \delta)^2 + \frac{t^2 + (v-t)^2}{t(v-t)}(e^\epsilon+\delta)(1-\delta) + (1-\delta)^2  }.
    \end{multline}
    If $v$ is even, then $t = v/2$ maximizes~\eqref{eq:F_LDP_t} to yield
    \begin{equation}\label{eq:F_LDP_1}
        1 + \left( \frac{e^\epsilon + 2\delta - 1}{e^\epsilon + 1} \right)^2.
    \end{equation}
    If $v = 2\alpha + 1$, $\alpha \in \mathbb{N}$, then $t = \alpha$ maximizes~\eqref{eq:F_LDP_t} to yield 
    \begin{equation}\label{eq:F_LDP_2}
        1 + \frac{(e^\epsilon+2\delta-1)^2} {(e^\epsilon+\delta)^2+ \frac{\alpha^2 + (\alpha+1)^2}{\alpha(\alpha+1)} (e^\epsilon+\delta)(1-\delta) + (1-\delta)^2 }.
    \end{equation}
    3) Suppose $Q$ has a column $q \in \{\delta,0\}^v$ and $t\geq 1$ elements of $q$ are $\delta$. Then,
    \begin{equation}
        F(Q) = 2 - \frac{(1-\delta)v}{v-\delta t}.
    \end{equation}
    Among all $t \geq 1$, $t=1$ maximizes the above to yield
    \begin{equation}\label{eq:F_LDP_3}
        1 + \frac{\delta(v-1)}{v-\delta}.
    \end{equation}

    By comparing~\eqref{eq:F_LDP_1},~\eqref{eq:F_LDP_2}, and~\eqref{eq:F_LDP_3}, we obtain a closed-form expression of $\sup_{Q \in \mathcal{Q}^{(\epsilon,\delta)}} F(Q)$.
    By substituting this into Lemma~\ref{lem:LAN_LB}, we have the desired lower bound of $\mathrm{PUT}^{\mathrm{LDP}}_{\mathrm{SR}}$.
    The conditions $\epsilon \geq \zeta(v,\delta)$ can be derived by solving the inequalities~\eqref{eq:F_LDP_1}$~\geq~$\eqref{eq:F_LDP_3} and~\eqref{eq:F_LDP_2}$~\geq~$\eqref{eq:F_LDP_3} with respect to $\epsilon$, respectively.
    
    For the $\gamma$-ML constraint, $F(Q) = 1$ if $Q$ has a zero column.
    Also, $F(Q) = 2 - \frac{(1-\delta)v}{v-\delta t}$ if $Q$ has a column $q \in \{e^\gamma -1 , 0\}^v$ and $t \geq 1$ elements of $q$ are $e^\gamma - 1$.
    Because $t=1$ maximizes $F(Q)$ to yield $1+\frac{(e^\gamma -1)(v-1)}{v-e^\gamma +1}$, Lemma~\ref{lem:LAN_LB} gives the desired result.
\end{IEEEproof}

Combining Propositions~\ref{prop:model_equiv_conv} and~\ref{prop:PUT_SR_LB}, we have the converse part of Theorem~\ref{thm:main}.

\section{Achievability} \label{sec:achiev}
In this section, we prove the achievability part of Theorem~\ref{thm:main}.
We aim to show that $\mathrm{PUT} \leq \mathrm{PUT}_{\mathrm{SR}}$ in Section~\ref{sec:compare_models_ach}.
To do so, we first construct optimal private estimation schemes with shared randomness that achieve $\mathrm{PUT}_{\mathrm{SR}}$.
Based on the structures of the optimal schemes with shared randomness, we construct private estimation schemes so that they resemble the optimal private estimation schemes with shared randomness asymptotically as the number of clients $n$ tends to infinity.

\subsection{Optimal schemes with shared randomness}\label{sec:opt_scheme}
In this subsection, we construct optimal schemes with shared randomness that achieve $\mathrm{PUT}_{\mathrm{SR}}$.
Some of our privacy mechanisms with shared randomness are closely related to the \emph{resolution of block design or regular and pairwise-balanced design (RPBD) mechanisms} proposed in \cite{park23_block,nam23_res_BD}.
For the estimator, we propose estimators that differ from  those in \cite{park23_block, nam23_res_BD}, because the previous estimators cannot be used directly for our privacy mechanisms with shared randomness.
By doing so, the proposed schemes with shared randomness are shown to achieve $\mathrm{PUT}_{\mathrm{SR}}$.

\subsubsection{Optimal privacy mechanisms}\label{sec:opt_mech}
First, we construct optimal privacy mechanisms with shared randomness.
Some of them are constructed based on the concept of a block design mechanism and an RPBD mechanism proposed in \cite{park23_block}, and resolutions of them \cite{nam23_res_BD}.
Here, we introduce such concepts with slight modifications.

\begin{definition}
    A hypergraph $G=(V,E)$, where $V$ is the set of the vertices and $E$ is the set of the edges, is called a $(v,b,r,k,\lambda)$-block design if $|V|=v$, $|E| = b$, and have the following symmetries:
    \begin{enumerate}
        \item Degree of each vertex is $r$ ($G$ is $r$-regular).
        \item Each edge contains $k$ vertices ($G$ is $k$-uniform).
        \item Each pair of vertices is contained in $\lambda$-number of edges ($G$ is $\lambda$-pairwise balanced).
    \end{enumerate}
    A hypergraph $G=(V,E)$ is called a $(v,b,r,\lambda)$-RPBD if $|V|=v$, $|E|=b$, and $G$ is $r$-regular and $\lambda$-pairwise balanced.
\end{definition}
\begin{remark}
    A block design of special interest in our work is a {\em  complete block design} (CBD).
    The $(v,k)$-CBD is the complete $k$-uniform hypergraph with $v$ vertices.
    It can be easily checked that the $(v,k)$-CBD is the block design with parameters
    \begin{equation}\label{eq:CBD}
        \left(v, \binom{v}{k}, \binom{v-1}{k-1}, k, \binom{v-2}{k-2} \right),
    \end{equation}
    with the convention that $\binom{v}{t} = 0$ for $t < 0$.
\end{remark}

\begin{definition}\label{def:incidence}
    Let $G=(V,E)$ be a hypergraph such that $V=\{v_1,\ldots,v_m\}$ and $E=\{e_1,\ldots,e_n\}$.
    The incidence matrix of $G$ is the matrix $A \in \{0,1\}^{m \times n}$ such that $A_{ij} = 1$ if $v_i \in e_j$ and $A_{ij}=0$ if $v_i \notin e_j$.
\end{definition}

\begin{definition}\label{def:BDmech}
    For $c,d \geq 0$, a stochastic matrix $Q$ of dimension $v \times b$ is called a $(c,d)$-valued $(v,b,r,k,\lambda)$-block design mechanism constructed by a block design $G$ if $G$ is a $(v,b,r,k,\lambda)$-block design and $Q$ can be constructed as follows:
    Let $A \in \{0,1\}^{v \times b}$ be an incidence matrix of $G$.
    Then, we get the matrix $B$ by applying the map $1 \mapsto c$ and $0 \mapsto d$ component-wisely on $A$.
    The stochastic matrix $Q$ is constructed as $Q = \frac{1}{cr + d(b-r)}B$.
    Similarly, a stochastic matrix $Q$ of dimension $v \times b$ is called a $(c,d)$-valued $(v,b,r,\lambda)$-RPBD mechanism constructed by an RPBD $G$ if $G$ is a $(v,b,r,\lambda)$-RPBD and $Q$ is constructed in the same way as above.
\end{definition}
% \begin{remark}
%     For a given block design (or RPBD) $G$, an incidence matrix of $G$ depends on the choice of the labeling on the vertices and the edges.
%     Though, all incidence matrices of $G$ are same up to row and column permutations.
%     Thus, the block design (or RPBD) mechanisms constructed by $G$ is defined up to equivalence.
%     % As we will see throughout this section, some of our proposed optimal schemes with shared randomness are resolutions of block design (or RPBD) scheme, and all equivalent form of them have the same worst-case MSE.
% \end{remark}

\begin{definition}
    A pair $(P_U,\tilde{Q})$ of a probability mass function $P_U \in \mathcal{P}(\mathcal{U})$ and a conditional probability mass function $\tilde{Q}:\mathcal{U} \times \mathcal{X} \rightarrow \mathcal{P}(\mathcal{Z})$ is called a resolution of a block design (or RPBD) mechanism $Q:\mathcal{X} \rightarrow \mathcal{P}(\mathcal{W})$ if $P_U \tilde{Q} : \mathcal{X} \rightarrow \mathcal{P}(\mathcal{U} \times \mathcal{Z})$ is equivalent to $Q$.
\end{definition}

\begin{figure*}
    \centering
    \includegraphics[width=.75\linewidth]{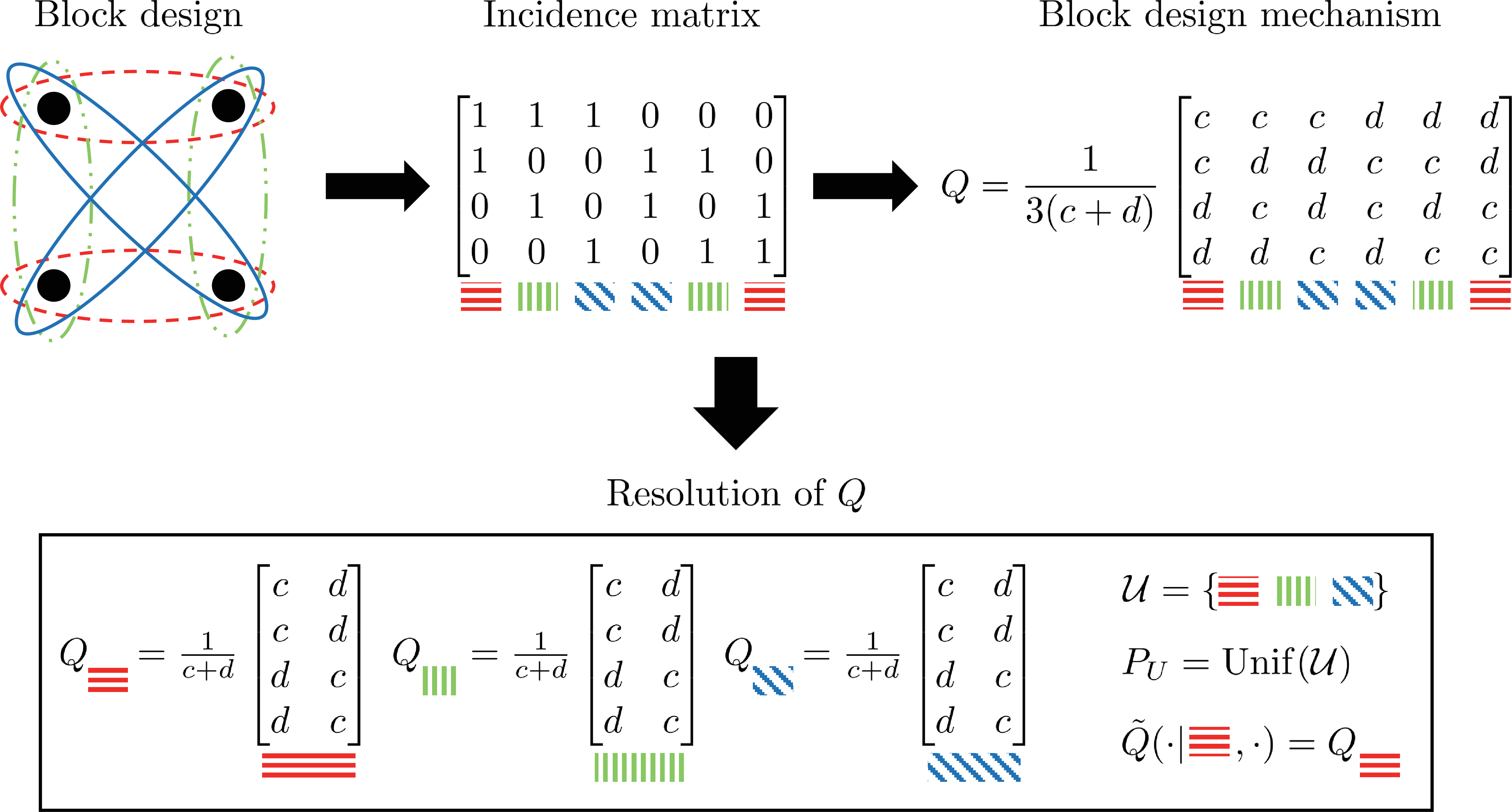}
    \caption{Schematic diagram showing the constructions of the block design mechanism constructed by $(4,2)$-CBD and its resolution.}
    \label{fig:res_ex}
\end{figure*}

\begin{example}\label{ex:res}
    We introduce an example of the detailed process of constructing a $(c,d)$-valued block design mechanism and its resolution, which is depicted in Fig.~\ref{fig:res_ex}.
    Let $G$ be the $(4,2)$-CBD, whose incidence matrix $A$ is
    \begin{equation}
        A = \begin{bmatrix}
        1 & 1 & 1 & 0 & 0 & 0 \\
        1 & 0 & 0 & 1 & 1 & 0 \\
        0 & 1 & 0 & 1 & 0 & 1 \\
        0 & 0 & 1 & 0 & 1 & 1 
        \end{bmatrix}.
    \end{equation}
    By applying the map $1 \mapsto c$ and $0 \mapsto d$ on $A$ component-wisely, we get,
    \begin{equation}
        B = \begin{bmatrix}
        c & c & c & d & d & d \\
        c & d & d & c & c & d \\
        d & c & d & c & d & c \\
        d & d & c & d & c & c 
        \end{bmatrix}.
    \end{equation}
    Then, by normalizing $B$, we get the block design mechanism $Q$ constructed by $G$,
    \begin{equation}
        Q = \frac{1}{3(c+d)}\begin{bmatrix}
        c & c & c & d & d & d \\
        c & d & d & c & c & d \\
        d & c & d & c & d & c \\
        d & d & c & d & c & c
        \end{bmatrix}.
    \end{equation}
    
    Now, let $Q^i$ be the $i$-th column of $Q$.
    Note that the columns of $Q$ can be partitioned into
    \begin{equation}
        C_1 = (Q^1, Q^6),\; C_2 = (Q^2, Q^5),\; C_3 = (Q^3,Q^4),
    \end{equation}
    and $Q^i + Q^{7-i} \propto \mathbf{1}$.
    Thus, we can get the stochastic matrices $Q_i = 3 C_i $ for each $i \in [3]$.
    Then, let $\mathcal{U} = [3]$, $P_U =\mathrm{Unif}(\mathcal{U})$, $\mathcal{Z} = [2]$, and $\tilde{Q}:\mathcal{U} \times \mathcal{X} \rightarrow \mathcal{P}(\mathcal{Z})$, $\tilde{Q}(\cdot|u,\cdot) = Q_u$ for each $u \in \mathcal{U}$.
    Clearly, $P_U \tilde{Q} \cong Q$.
\end{example}

Now, we construct the optimal privacy mechanisms with shared randomness.
Throughout this section, we treat conditional probability mass functions as stochastic matrices without loss of generality.
The constructions of the optimal mechanisms with shared randomness are closely related to the optimal solutions of $\sup_{Q \in \mathcal{Q}}F(Q)$, which are in the proof of Proposition~\ref{prop:PUT_SR_LB}.
We propose four privacy mechanisms with shared randomness in order, the first three are for the LDP constraint and the last one is for the ML constraint.
As in Example~\ref{ex:res}, optimal privacy mechanisms are derived by partitioning the columns of stochastic matrices into the \emph{dual pairs}.
\begin{definition}
    Let $Q$ be a stochastic matrix of a finite dimension, and $Q^i$ be the $i$-th column of $Q$.
    We call a pair of columns $(Q^i,Q^j)$ is a dual pair if $i \neq j$ and $Q^i + Q^j \propto \mathbf{1}$.
\end{definition}

\textbf{Case 1.} Assume $v$ is even and $\epsilon \geq \zeta(v,\delta)$.
% In this case, the optimal solution of $\sup_{Q \in \mathcal{Q}^{(\epsilon,\delta)}}F(Q)$ is the stochastic matrix of dimension $v\times 2$ which has a column $q \in \left\{\frac{e^\epsilon+\delta}{e^\epsilon+1}, \frac{1-\delta}{e^\epsilon+1} \right\}^v$, and $v/2$ elements of $q$ are $\frac{e^\epsilon+\delta}{e^\epsilon+1}$, as in the proof of Proposition~\ref{prop:PUT_SR_LB}.
Let $(c,d) = \left(\frac{e^\epsilon+\delta}{e^\epsilon+1}, \frac{1-\delta}{e^\epsilon+1} \right)$, and $G$ be the $(v,v/2)$-CBD.
Note that the $(v,v/2)$-CBD is a $(v,b,r,v/2,\lambda)$-block design, where
\begin{equation}
    (b,r,\lambda)= \left( \binom{v}{v/2}, \frac{1}{2}\binom{v}{v/2}, \binom{v-2}{v/2-2} \right).
\end{equation}
Now, we construct $Q$ as the $(c,d)$-valued block design mechanism constructed by $(v,v/2)$-CBD.
Then, the columns of $Q$ can be partitioned into the dual pairs $C_1,\ldots,C_{b/2}$ because for any given edge $e$ of $(v,v/2)$-CBD, there exists a unique edge $e'$ such that $|e \cup e'| = v$.
Using this fact, we construct a privacy mechanism with shared randomness $(P_U,\tilde{Q})$ as 
\begin{equation}\label{eq:opt_mech_const1}
    \mathcal{U} = \left[ b/2 \right],\quad P_U = \mathrm{Unif}(\mathcal{U}),
\end{equation}
\begin{equation}
     \mathcal{Z} = [2],\quad \tilde{Q}:\mathcal{U} \times \mathcal{X} \rightarrow \mathcal{P}(\mathcal{Z}),
\end{equation}
\begin{equation}\label{eq:opt_mech_const2}
    \forall u \in \mathcal{U},\; \tilde{Q}(\cdot | u, \cdot ) = \frac{cr+d(b-r)}{c+d}C_u = \frac{b}{2}C_u.
\end{equation}
By construction, $(P_U,\tilde{Q})$ is a resolution of $Q$.
Also, it can be easily seen that $(P_U,\tilde{Q}) \in \tilde{\mathcal{Q}}^{(\epsilon,\delta)}$ because $|\mathcal{Z}|=2$ and $e^\epsilon d + \delta \geq c$.

\textbf{Case 2.} Assume $v = 2\alpha + 1$, $\alpha \in \mathbb{N}$, and $\epsilon \geq \zeta(\alpha,\delta)$.
% In this case, the optimal solution of $\sup_{Q \in \mathcal{Q}^{(\epsilon,\delta)}}F(Q)$ is the stochastic matrix of dimension $v \times 2$ which has a column $q \in \left\{\frac{e^\epsilon+\delta}{e^\epsilon+1}, \frac{1-\delta}{e^\epsilon+1} \right\}^v$, and $\alpha$ elements of $q$ are $\frac{e^\epsilon+\delta}{e^\epsilon+1}$.
Let $(c,d) = \left(\frac{e^\epsilon+\delta}{e^\epsilon+1}, \frac{1-\delta}{e^\epsilon+1}\right)$, $G_1=(\mathcal{X},E_1)$ be the $(v,\alpha)$-CBD, $G_2 = (\mathcal{X},E_2)$ be the $(v,\alpha+1)$-CBD, and $G = (\mathcal{X},E_1 \cup E_2)$.
Then, $G$ is a $(v,b,r,\lambda)$-RPBD, where
\begin{equation}\label{eq:param_concat}
    (b,r,\lambda) =
    \left( 2\binom{v}{\alpha}, \binom{v}{\alpha}, \binom{v-1}{\alpha-1} \right).
\end{equation}
Now, we construct $Q$ as the RPBD mechanism constructed by $G$.
Then, the columns of $Q$ can be partitioned into dual pairs $C_1,\ldots,C_{b/2}$ because for any edge $e$ of $G_1$, there exists a unique edge $e'$ of $G_2$ such that $|e \cup e'| = v$.
Then, $(P_U,\tilde{Q})$ is constructed as in~\eqref{eq:opt_mech_const1}--\eqref{eq:opt_mech_const2}.
Similar to Case 1, $(P_U,\tilde{Q}) \in \tilde{\mathcal{Q}}^{(\epsilon,\delta)}$, and it is a resolution of $Q$.

\textbf{Case 3.} Assume that $\epsilon<\zeta(v,\delta)$.
% In this case, the optimal solution of $\sup_{Q \in \mathcal{Q}^{(\epsilon,\delta)}}F(Q)$ is the stochastic matrix of dimension $v \times 2$ which has a column $q \in \{\delta,0\}^v$, and only one element of $q$ is $\delta$.
We construct $Q$ as
\begin{equation}
    Q = \frac{1}{v}\left(\delta I_{(v\times v)},\; \mathbf{1}_{(v\times v)} - \delta I_{(v\times v)} \right).
\end{equation}
Let $Q^i$ be the $i$-th column of $Q$, $i \in [2v]$.
Then, the columns of $Q$ can be partitioned into the dual pairs $C_1,\ldots,C_v$, where $C_i = (Q^i,Q^{i+v})$, $i \in [v]$.
Accordingly, we construct a privacy mechanism with shared randomness $(P_U,\tilde{Q})$ as
\begin{equation}\label{eq:opt_mech_const_3}
    \mathcal{U} = [v], \quad P_U = \mathrm{Unif}(\mathcal{U}), \quad \mathcal{Z} = [2],
\end{equation}
\begin{equation}\label{eq:opt_mech_const_4}
    \tilde{Q}:\mathcal{U} \times \mathcal{X} \rightarrow \mathcal{P}(\mathcal{Z}),\quad \forall u \in \mathcal{U}, \; \tilde{Q}(\cdot|u,\cdot) = vC_i.
\end{equation}
Then, $P_U \tilde{Q} \cong Q$ and $(P_U,\tilde{Q}) \in \tilde{\mathcal{Q}}^{(\epsilon,\delta)}$.

\textbf{Case 4.} 
In this case, we consider the $\gamma$-ML constraint.
% Note that the optimal solution of $\sup_{Q \in \mathcal{Q}^\gamma} F(Q)$ is the stochastic matrix of dimension $v \times 2$ which has a column $q \in \{e^\gamma -1,0\}^v$, and only one element of $q$ is $e^\gamma-1$.
We construct $Q$ as
\begin{equation}
    Q = \frac{1}{v}\left((e^\gamma-1) I_{(v\times v)},\; \mathbf{1}_{(v\times v)} - (e^\gamma-1) I_{(v\times v)} \right).
\end{equation}
Similar to Case 3, the columns of $Q$ can be partitioned into the dual pairs $C_1,\ldots,C_v$, where $C_i = (Q^i,Q^{i+v})$, $i \in [v]$.
A privacy mechanism with shared randomness $(P_U,\tilde{Q})$ is constructed as in~\eqref{eq:opt_mech_const_3} and~\eqref{eq:opt_mech_const_4}.
Then, $P_U \tilde{Q} \cong Q$ and $(P_U,\tilde{Q}) \in \tilde{\mathcal{Q}}^{\gamma}$.

\subsubsection{Optimal estimators}
Note that all the four privacy mechanisms with shared randomness $(P_U,\tilde{Q})$ that we have constructed in Section~\ref{sec:opt_mech} are derived by some pre-designed stochastic matrices $Q$ such that $P_U \tilde{Q} \cong Q$.
The proposed estimator $\tilde{\theta}_n:\mathcal{U}^n \times \mathcal{Z}^n \rightarrow \mathbb{R}^v$ is constructed based on such $Q$.
Without loss of generality, we treat the stochastic matrix $Q$ as the conditional probability mass function $Q:\mathcal{X} \rightarrow \mathcal{P}(\mathcal{W})$, $\mathcal{W} = \mathcal{U} \times \mathcal{Z}$, and denote $W=(U,Z)$ as the random variable sampled from $Q$.
For a given $Q$, we construct the estimator $\tilde{\theta}_n$ as follows:
We first design the auxiliary estimator $\eta_n:\mathcal{W}^n \rightarrow \mathbb{R}^v$.
Let $\eta:\mathcal{W} \rightarrow \mathbb{R}^v$, whose components are the normalized likelihoods,
\begin{equation}\label{eq:eta_x}
    \forall x \in \mathcal{X},\quad \eta_x(w) = \frac{Q(w|x)}{\sum\limits_{x' \in \mathcal{X}}Q(w|x')}.
\end{equation}
Then, $\eta_n:\mathcal{W}^n \rightarrow \mathbb{R}^v$ is set to the average of $\eta(w_i)$'s,
\begin{equation}\label{eq:def_eta}
    \eta_n(w^n) = \frac{1}{n} \sum\limits_{i=1}^n \eta(w_i).
\end{equation}
Note that $\mathbb{E}[\eta_n(W^n)] = \mathbb{E}[\eta(W)]$ because $W_1,\ldots,W_n$ are i.i.d.
In the following lemmas, we show that there exist real constants $c_1\neq 0, c_2$ such that $\mathbb{E}[\eta(W)] = c_1 \theta + c_2 \mathbf{1}$, if we use the $Q$ constructed in Section~\ref{sec:opt_mech}.
Finally, we construct $\tilde{\theta}_n$ as an unbiased version of $\eta_n$,
\begin{equation}\label{eq:opt_est}
    \tilde{\theta}_n (w^n) = \frac{1}{c_1}(\eta_n(w^n) - c_2 \mathbf{1}).
\end{equation}
The proofs of the following lemmas are in Appendix.
\begin{lemma}\label{lem:eta_calc_1}
    Let $v$ be even and $Q:\mathcal{X} \rightarrow \mathcal{P}(\mathcal{W})$ be the $(c,d)$-valued block design mechanism constructed by the $(v,v/2)$-CBD.
    Also, for each $x \in \mathcal{X}$, define the sets
    \begin{align}
        A^1_x &= \left\{ w \in \mathcal{W}: Q(w|x) = \frac{2c}{\binom{v}{v/2} (c+d)} \right\},\\
        A^2_x& = \left\{ w \in \mathcal{W}: Q(w|x) = \frac{2d}{\binom{v}{v/2} (c+d)} \right\}.
    \end{align}
    Then, for all $x \in \mathcal{X}$, 
    \begin{equation}\label{eq:eta_x_1}
        \eta_x(w) = \frac{2}{v(c+d)}\left( c\mathbbm{1}(w \in A^1_x) + d\mathbbm{1}(w \in A^2_x) \right),
    \end{equation}
    and
    \begin{multline}\label{eq:eta_first_coeff_1}
        \mathbb{E}[\eta_x(W)] = \frac{(c-d)^2}{(v-1)(c+d)^2} \theta_x
        \\ + \frac{v(c+d)^2-2(c^2+d^2)}{v(v-1)(c+d)^2}.
    \end{multline}
\end{lemma}

\begin{lemma}\label{lem:eta_calc_2}
    Let $v = 2\alpha + 1$, $\alpha \in \mathbb{N}$, $G_1 = (\mathcal{X},E_1)$ be the $(v,\alpha)$-CBD, $G_2 = (\mathcal{X},E_2)$ be the $(v,\alpha+1)$-CBD, $G = (\mathcal{X},E_1 \cup E_2)$, 
    and $Q:\mathcal{X} \rightarrow \mathcal{P}(\mathcal{W})$ be the $(c,d)$-valued RPBD mechanism constructed by $G$.
    Also, define the sets
    \begin{equation}
        H=
        \left\{w \!\in\! \mathcal{W} \!:\! \sum\limits_{x \in \mathcal{X}}\! \mathbbm{1}\left(\! Q(w|x)\!=\!\frac{c}{\binom{v}{\alpha}(c\!+\!d)} \right) \!=\! \alpha  \right\},
    \end{equation}
    and for each $x \in \mathcal{X}$,
    \begin{align}
        G_x=\left\{w \in \mathcal{W}: Q(w|x) = \frac{c}{\binom{v}{\alpha}(c+d)}\right\},
    \end{align}
    and 
    \begin{align}
        A_x^1 = G_x \cap H,&\quad A_x^2 = G_x \cap H^\complement,
        \\ A_x^3 = G_x^\complement \cap H, &\quad A_x^4 = G_x^\complement \cap H^\complement.
    \end{align}
    Then, for all $x \in \mathcal{X}$, 
    \begin{multline}\label{eq:eta_x_2}
        \eta_x(w) = \frac{c \mathbbm{1}(w \in A^1_x) + d \mathbbm{1}(w \in A^3_x)}{\alpha c + (\alpha+1)d}
        \\ + \frac{c \mathbbm{1}(w \in A^2_x) + d \mathbbm{1}(w \in A^4_x)}{\alpha d + (\alpha+1)c},
    \end{multline}
    and
    \begin{multline}\label{eq:eta_first_coeff_2}
        \mathbb{E}[\eta_x(W)] = \frac{(c-d)^2(\alpha+1)}{2((\alpha+1)c +\alpha d)(\alpha c + (\alpha+1)d )} \theta_x
        \\ + \frac{(2\alpha+1)((c+d)^2\alpha + 2cd)-(c-d)^2}{2(2\alpha+1)((\alpha+1)c +\alpha d)(\alpha c + (\alpha+1)d )} .
    \end{multline}
\end{lemma}

\begin{lemma}\label{lem:eta_calc_3}
    For $c> 0$, let the stochastic matrix $Q:\mathcal{X} \rightarrow \mathcal{P}(\mathcal{W})$ be
    \begin{equation}\label{eq:Q_diag_shape}
        Q = \frac{1}{v} \left( c I_{(v\times v)},\; \mathbf{1}_{(v\times v)} - c I_{(v\times v)} \right).
    \end{equation}
    Also, for each $x \in \mathcal{X}$, define the sets
    \begin{align}
        A^1_x &= \{w \in \mathcal{W}: Q(w|x) = c/v\},\\
        A^2_x &= \{w \in \mathcal{W}: Q(w|x) = 1/v\},\\
        A^3_x &= \{w \in \mathcal{W}: Q(w|x) = (1-c)/v\}.
    \end{align}
    Then, for all $x \in \mathcal{X}$, 
    \begin{multline}\label{eq:eta_x_3}
        \eta_x(w) = \mathbbm{1}(w \in A^1_x)
        \\ + \frac{1}{v-c} \left( \mathbbm{1}(w \in A^2_x) + (1-c) \mathbbm{1}(w \in A^3_x) \right),
    \end{multline}
    and
    \begin{equation}\label{eq:eta_first_coeff_3}
        \mathbb{E}[\eta_x(W)] = \frac{c}{v-c} \theta_x + \frac{v-2c}{v(v-c)}.
    \end{equation}
\end{lemma}

\subsubsection{Error analysis}\label{sec:Error_opt_scheme}
The estimation error of the proposed schemes $(P_U,\tilde{Q},\tilde{\theta}_n)$ that we have constructed in this section are calculated in the same manner.
We first introduce the calculation steps before the detailed calculation.
Because $\tilde{\theta}_n$ in~\eqref{eq:opt_est} is an unbiased estimator, the MSE of the proposed scheme is  \begin{align}
    \mathbb{E}\left[\norm{\theta-\tilde{\theta}_n(W^n)}_2^2 \right] &= \sum\limits_{x \in \mathcal{X}} \mathrm{Var} \left((\tilde{\theta}_n)_x(W^n)\right)
    % \\&= \frac{1}{c_1^2} \sum\limits_{x \in \mathcal{X}} \mathrm{Var}((\eta_n)_x(W^n))
    \\& = \frac{1}{n c_1^2} \sum\limits_{x \in \mathcal{X}} \mathrm{Var}(\eta_x(W)). \label{eq:MSE_concave}
\end{align}
Note that $\eta_x$'s in~\eqref{eq:eta_x_1},~\eqref{eq:eta_x_2}, and~\eqref{eq:eta_x_3} have the form of
\begin{equation}
    \eta_x(w) = \sum\limits_i \lambda_i \mathbbm{1}(w \in A_x^i),
\end{equation}
for some disjoint sets $A_x^i$'s and normalization factors $\lambda_i$'s.
Thus, we have
\begin{equation}\label{eq:var_eta_x}
    \mathrm{Var}(\eta_x(W)) = \sum\limits_i \lambda_i^2 P_x^i(1-P_x^i) - \sum\limits_{i \neq j} \lambda_i \lambda_j P_x^i P_x^j,
\end{equation}
where $P_x^i = \mathrm{Pr}(W \in A_x^i)$.
Then, we will show that $\mathrm{Var}(\eta_x(W))$ is a concave function of $\theta_x$ for all $x \in \mathcal{X}$ in Lemma~\ref{lem:var_eta_conc}.
Because the MSE in~\eqref{eq:MSE_concave} is the sum of the component-wise concave functions, the MSE is the concave function of $\theta$.
Together with the fact that~\eqref{eq:MSE_concave} does not vary under any permutation on $\theta$, the worst-case MSE is achieved by $\theta = \mathbf{1}/v$, i.e.,
\begin{align}
    R_{n,v}(P_U,\tilde{Q},\tilde{\theta}_n)
    = \mathbb{E}_{X_1,\ldots,X_n \sim \frac{\mathbf{1}}{v}}\left[\norm{\frac{\mathbf{1}}{v} -\tilde{\theta}_n(W^n)}_2^2 \right].\label{eq:MSE_unif}
\end{align}
Finally, we calculate~\eqref{eq:var_eta_x} for $\theta = \mathbf{1}/v$ and substitute the results into~\eqref{eq:MSE_concave}.
By doing so, it can be shown that the worst-case MSEs of the four schemes with shared randomness that we have constructed in this section are equal to each of the lower bounds of $\mathrm{PUT}_{\mathrm{SR}}/n$.

As we mentioned above, we first check that $\mathrm{Var}(\eta_x)$ is a concave second order polynomial in $\theta_x$ for all $x \in \mathcal{X}$, and then calculate the worst-case MSE by letting $\theta = \mathbf{1}/v$. 
The proofs of the following lemma and proposition can be found in the supplementary material.
\begin{lemma}\label{lem:var_eta_conc}
    For any private estimation scheme with shared randomness constructed in Section~\ref{sec:opt_scheme}, $\mathrm{Var}(\eta_x(W))$ is a concave second order polynomial in $\theta_x$.
\end{lemma}

\begin{proposition}\label{prop:PUT_SR_achiev}
    The worst-case MSEs of each of the four private estimation schemes with shared randomness constructed in Section~\ref{sec:opt_scheme} are equal to each of the lower bounds of $\mathrm{PUT}_{\mathrm{SR}}/n$ in Proposition~\ref{prop:PUT_SR_LB}.
\end{proposition}

\subsection{Comparing models: Achievability} \label{sec:compare_models_ach}
In this subsection, we prove $\mathrm{PUT} \leq \mathrm{PUT}_{\mathrm{SR}}$ as the last step of the proof of Theorem~\ref{thm:main}.
For showing $\mathrm{PUT} \leq \mathrm{PUT}_{\mathrm{SR}}$, we construct the private estimation schemes $(Q_1,\ldots,Q_n,\hat{P}_n)$ that asymptotically resemble the optimal private estimation schemes with shared randomness $(P_U,\tilde{Q},\tilde{\theta}_n)$ as the number of clients $n$ tends to infinity.

\begin{proposition}\label{prop:equiv_model_achiev}
We have that
    \begin{equation}
        \mathrm{PUT} \leq \mathrm{PUT}_{\mathrm{SR}}.
    \end{equation}    
\end{proposition}

\begin{IEEEproof}
    Let $(P_U,\tilde{Q},\tilde{\theta}_n)$ be the optimal scheme with shared randomness constructed in Section~\ref{sec:opt_scheme}.
    Note that for any optimal scheme with shared randomness in Section~\ref{sec:opt_scheme}, $\mathcal{U} = [C]$ for some constant $C \in \mathbb{N}$, $P_U = \mathrm{Unif}(\mathcal{U})$, 
    \begin{equation}
        \tilde{\theta}_n(u^n,z^n) = \frac{1}{c_1}\left( \frac{1}{n}\sum\limits_{i=1}^n \eta(u_i,z_i)  - c_2 \mathbf{1}\right),
    \end{equation}
    for some constants $c_1,c_2$, and $\tilde{\theta}_n$ is unbiased.
    Thus, we have
    \begin{equation}\label{eq:last_unbiased}
        \mathbb{E}[\tilde{\theta}_n(U^n,Z^n)] = \frac{1}{c_1} \left( \mathbb{E}[\eta(U,Z)] - c_2 \mathbf{1} \right) = \theta,
    \end{equation}
    and
    \begin{equation}\label{eq:last_var_UZ}
        \mathbb{E}\left[ \norm{\theta - \tilde{\theta}_n(U^n,Z^n)}_2^2 \right] = \frac{1}{nc_1^2} \sum\limits_{x \in \mathcal{X}} \mathrm{Var}(\eta_x(U,Z)),
    \end{equation}
    because $(U_1,Z_1),\ldots,(U_n,Z_n)$ are i.i.d.
    
    Without loss of generality, we assume $n > C$.
    First, let the private estimation scheme $(Q_1,\ldots,Q_n)$ be
    \begin{equation}
        \mathcal{Y}_i = \mathcal{Z}_{g_n(i)} ,\quad Q_i = \tilde{Q}(\cdot|g_n(i),\cdot),
    \end{equation}
    where $g_n:[n] \rightarrow \mathcal{U}$, $g_n(i) = ((i-1) \mod C) + 1$.
    Clearly, $Q_i \in \mathcal{Q}$ for all $i \in [n]$.
    Then, we define the estimator $\hat{\theta}_n:\mathcal{Y}^n \rightarrow \mathbb{R}^v$ as
    \begin{multline}\label{eq:last_hat_thet}
        \hat{\theta}_n(y^n) =\frac{1}{c_1}\times
        \\ \left(\frac{1}{\lfloor n/C \rfloor} \sum\limits_{i=1}^{\lfloor n/C \rfloor} \frac{1}{C} \sum\limits_{j=1}^C  \eta(j,y_{(i-1)C + j}) - c_2 \mathbf{1}\right). 
    \end{multline}
    By the law of total expectation, we have
    \begin{align}
        \mathbb{E}[\eta(U,Z)] = \frac{1}{C}\sum\limits_{j=1}^C \mathbb{E}[\eta(j,Z)|U=j]
         = \frac{1}{C}\sum\limits_{j=1}^C \mathbb{E}[\eta(j,Y_j)],\label{eq:exp_eta}
    \end{align}
    where the last equation follows from the fact that for any given $U=j$, $Z$ and $Y_j$ follow the same distribution.
    Also, note that $(Y_1,\ldots,Y_C)$ and $(Y_{(i-1)C+1},\ldots,Y_{iC})$ are independent and follow the same distribution for all $i \in [\lfloor n/C \rfloor]$.
    Together with~\eqref{eq:last_unbiased} ,\eqref{eq:last_hat_thet}, and~\eqref{eq:exp_eta}, we have
    \begin{equation}
        \mathbb{E}[\hat{\theta}_n(Y^n)] = \frac{1}{c_1} \left( \frac{1}{C}\sum\limits_{j=1}^C \mathbb{E}[\eta(j,Y_j)] - c_2 \mathbf{1} \right) = \theta.
    \end{equation}
    Because $\hat{\theta}_n$ is unbiased, we obtain
    \begin{align}
        \mathbb{E}&\left[ \norm{\theta - \hat{\theta}_n(Y^n)}_2^2 \right]\nonumber
        \\&= \frac{1}{c_1^2 \lfloor n/C \rfloor  C^2} \sum\limits_{x \in \mathcal{X}}\sum\limits_{j=1}^C \mathrm{Var}(\eta_x(j,Y_j))
        \\&= \frac{1}{c_1^2 \lfloor n/C \rfloor  C} \sum\limits_{x \in \mathcal{X}}\mathbb{E}\left[\mathrm{Var}(\eta_x(U,Z)|U)\right] \label{eq:last_pf_1}
        \\& \leq \frac{1}{(n-C)c_1^2} \sum\limits_{x \in \mathcal{X}}\mathrm{Var}(\eta_x(U,Z)),\label{eq:last_pf_2}
    \end{align}
    where~\eqref{eq:last_pf_1} follows from the fact that for any given $U=j$, $Z$ and $Y_j$ follow the same distribution, and the last inequality is from the law of total variance.
    Thus,~\eqref{eq:last_var_UZ} and~\eqref{eq:last_pf_2} yield
    \begin{align}
        \!\!\!\mathbb{E}\left[ \norm{\theta - \hat{\theta}_n(Y^n)}_2^2 \right]
         \!\leq\! \frac{n}{n-C}\mathbb{E}\left[ \norm{\theta - \tilde{\theta}_n(U^n,Z^n)}_2^2\right].\!
    \end{align}
    By taking the supremum over $\theta \in \mathcal{P}([v])$ on both sides and applying Proposition~\ref{prop:PUT_SR_achiev}, we have
    \begin{equation}
        \!\!\!\mathrm{PUT}_{n} \leq R_{n,v}(Q_1,\ldots,Q_n,\hat{\theta}_n) \leq \frac{1}{n-C} \mathrm{PUT}_{\mathrm{SR}}.
    \end{equation}
    Finally, we obtain the desired result by multiplying $n$ and taking $\liminf_{n\rightarrow\infty}$ on both sides.
\end{IEEEproof}

Combining Propositions~\ref{prop:model_equiv_conv},~\ref{prop:PUT_SR_LB},~\ref{prop:PUT_SR_achiev} and~\ref{prop:equiv_model_achiev}, we complete the proof of Theorem~\ref{thm:main}.

\section{Conclusion}\label{sec:conc}
In this paper, we completely characterized the PUTs for discrete distribution estimation under the $(\epsilon,\delta)$-LDP or $\gamma$-ML privacy constraints, together with the one-bit communication constraint.
For the converse part, we exploited the local asymptotic normality property as in \cite{ye17_opt_PUT_l2}, and found tight lower bounds by characterizing all the extreme points of the set of privacy mechanisms.
For the achievability part, we presented concrete schemes that achieve the optimal PUTs with the idea of resolutions of block design schemes \cite{park23_block,nam23_res_BD}.

One avenue for future investigation would be to characterize the PUCT under the $b$-bit communication constraint for arbitrary $b>1$.
For the converse part of such an endeavor, Proposition~\ref{prop:model_equiv_conv} and a variant of Lemma~\ref{lem:LAN_LB} still hold.
However, the full characterization of the extreme points of the set of privacy mechanisms is still not known (cf. \cite{holohan17_LDP_ex_pts}).
If we can obtain a complete characterization of the extreme points, it would be possible to derive lower bounds in a similar way as in the proof of Proposition~\ref{prop:PUT_SR_LB}.

\appendix

Here, we prove Lemmas \ref{lem:eta_calc_1},~\ref{lem:eta_calc_2}, and~\ref{lem:eta_calc_3} in Section~\ref{sec:opt_scheme}.
The proofs are based on calculations related to the combinatorial structures of the proposed scheme, which are similar to the calculations in \cite[Sec. III]{Ye18_SS}.
As we mentioned after~\eqref{eq:var_eta_x}, we denote $P_x^i = \mathrm{Pr}(W \in A_x^i)$.

\subsection{Proof of Lemma~\ref{lem:eta_calc_1}} \label{app:lem_1}
\begin{IEEEproof}
Let $k = v/2$.
Note that $Q$ is the $(c,d)$-valued block design mechanism constructed by $(v,k)$-CBD, and the $(v,k)$-CBD has the parameters $(v,b,r,k,\lambda)$ in~\eqref{eq:CBD}.
By Definition~\ref{def:BDmech},
\begin{equation}
    Q = \frac{1}{cr + d(b-r)} B,
\end{equation}
for some $\{c,d\}$-valued matrix $B$.
Then, we have
\begin{equation}
    cr + d(b-r) = \binom{v}{k} \frac{c+d}{   2}.
\end{equation}
Thus,~\eqref{eq:eta_x_1} directly follows from~\eqref{eq:eta_x}.
Note that $\{A_x^1,A_x^2\}$ is a partition of $\mathcal{W}$ and $P_x^2 = 1-P_x^1$.
Now, it remains to calculate $\mathbb{E}[\eta_x(W)]$.
Let
\begin{equation}
    \lambda_1 = \frac{2c}{v(c+d)},\; \lambda_2 = \frac{2d}{v(c+d)}.
\end{equation}
Then,
\begin{align}
    \mathbb{E}[\eta_x(W)] &= \lambda_1 P_x^1 + \lambda_2 P_x^2 = (\lambda_1 - \lambda_2) P_x^1 + \lambda_2 \nonumber\\
     &= \frac{2(c-d)}{v(c+d)} P_x^1 + \frac{2d}{v(c+d)}.\label{eq:calc_eta_1}
\end{align}
The calculation of $P_x^1$ is based on the combinatorial structure of $(v,k)$-CBD.
By Definition~\ref{def:BDmech}, $A_x^1$ corresponds to the set of edges of $(v,k)$-CBD containing a vertex which corresponds to $x \in \mathcal{X}$.
Thus, we have
\begin{align}\label{eq:calc_px_1}
    P_x^1 &= \frac{2 \left( \binom{v-1}{k-1}c \theta_x + \left(\binom{v-2}{k-2} c + \binom{v-2}{k-1} d \right)(1-\theta_x)  \right)}{\binom{v}{k}(c+d)} 
    % \\& = \frac{2\left( \binom{v-2}{k-1}(c-d) \theta_x + \binom{v-2}{k-2} c + \binom{v-2}{k-1} d  \right)}{\binom{v}{k}(c+d)} 
    \\& = \frac{v(c-d)}{2(v-1)(c+d)}\theta_x + \frac{(v-2)c + vd}{2(v-1)(c+d)}.\label{eq:calc_px_2}
\end{align}
By substituting the above into~\eqref{eq:calc_eta_1}, we have~\eqref{eq:eta_first_coeff_1}.
\end{IEEEproof}

\subsection{Proof of Lemma~\ref{lem:eta_calc_2}}\label{app:lem_2}
\begin{IEEEproof}
Let $v= 2\alpha + 1$, $\alpha \in \mathbb{N}$.
Note that $Q$ is a $(c,d)$-valued RPBD mechanism constructed by the RPBD with parameters $(v,b,r,\lambda)$ that equal to~\eqref{eq:param_concat}, as in Case 2 of Section~\ref{sec:opt_mech}.
By definition~\ref{def:BDmech}, 
\begin{equation}
    Q = \frac{1}{cr + d(b-r)}B,
\end{equation}
for some $\{c,d\}$-valued matrix $B$, and
\begin{equation}
    cr + d(b-r) = \binom{v}{\alpha} (c+d).
\end{equation}
Thus,~\eqref{eq:eta_x_2} directly follows from~\eqref{eq:eta_x}.
Now, let
\begin{align}
    \lambda_1 = \frac{c}{\alpha c + (\alpha+1)d},&\quad \lambda_2 = \frac{c}{(\alpha+1) c + \alpha d},
    \\\lambda_3 = \frac{d}{\alpha c + (\alpha+1)d},&\quad
    \lambda_4 = \frac{d}{(\alpha+1) c + \alpha d}.
\end{align}
Then,
\begin{equation}\label{eq:eta_x_2_calc}
    \mathbb{E}[\eta_x(W)] = \sum\limits_{i=1}^4 \lambda_i P_x^i.
\end{equation}
The calculations for $P_x^i$'s are similar to~\eqref{eq:calc_px_1}--\eqref{eq:calc_px_2}.
Using the fact that $A_x^1$ and $A_x^3$ correspond to $(v,\alpha)$-CBD, we have
\begin{align}
    P_x^1 &= \frac{\left( \binom{2\alpha}{\alpha-1}c \theta_x + \left(\binom{2\alpha-1}{\alpha-2} c + \binom{2\alpha-1}{\alpha-1} d \right)(1-\theta_x)  \right)}{\binom{2\alpha + 1}{\alpha}(c+d)} 
    \\& = \frac{(\alpha+1)(c-d)}{2(2\alpha+1)(c+d)} \theta_x + \frac{(\alpha-1)c + (\alpha+1)d}{2(2\alpha+1)(c+d)},
\end{align}
\begin{align}
    P_x^3 &= \frac{\left( \binom{2\alpha}{\alpha}d \theta_x + \left(\binom{2\alpha-1}{\alpha} d + \binom{2\alpha-1}{\alpha-1} c \right)(1-\theta_x)  \right)}{\binom{2\alpha + 1}{\alpha}(c+d)} 
    \\& = -\frac{(\alpha+1)(c-d)}{2(2\alpha+1)(c+d)} \theta_x + \frac{\alpha+1}{2(2\alpha+1)} .
\end{align}
Similarly, $A_x^2$ and $A_x^4$ correspond to $(v,\alpha+1)$-CBD.
Thus, we have
\begin{align}
    P_x^2
    % &= \frac{\left( \binom{2\alpha}{\alpha}c \theta_x + \left(\binom{2\alpha-1}{\alpha-1} c + \binom{2\alpha-1}{\alpha} d \right)(1-\theta_x)  \right)}{\binom{2\alpha + 1}{\alpha}(c+d)} 
    % \\&
    &= \frac{(\alpha+1)(c-d)}{2(2\alpha+1)(c+d)} \theta_x + \frac{\alpha+1}{2(2\alpha+1)},
    \\ P_x^4 &
    % &= \frac{\left( \binom{2\alpha}{\alpha+1}d \theta_x + \left(\binom{2\alpha-1}{\alpha+1} d + \binom{2\alpha-1}{\alpha} c \right)(1-\theta_x)  \right)}{\binom{2\alpha + 1}{\alpha}(c+d)} 
    % \\&
    = -\frac{(\alpha+1)(c-d)}{2(2\alpha+1)(c+d)} \theta_x + \frac{(\alpha-1)d+(\alpha+1)c}{2(2\alpha+1)(c+d)} .
\end{align}
By plugging the $\lambda_i$'s and $P_x^i$'s into~\eqref{eq:eta_x_2_calc}, we have~\eqref{eq:eta_first_coeff_2}.
\end{IEEEproof}

\subsection{Proof of Lemma~\ref{lem:eta_calc_3}}\label{app:lem_3}
\begin{IEEEproof}
It is easy to check~\eqref{eq:eta_x_3} from~\eqref{eq:Q_diag_shape}.
Let
\begin{equation}
    \lambda_1 = 1, \; \lambda_2 = \frac{1}{v-c},\; \lambda_3 = \frac{1-c}{v-c}.
\end{equation}
After simple calculations, we have $P_x^1 = c\theta_x / v$,
\begin{align}
    P_x^2 &= \frac{v-1}{v}\theta_x + \frac{v-1-c}{v} (1-\theta_x)
    \\& = \frac{c}{v} \theta_x + \frac{v-1-c}{v},
\end{align}
and
\begin{equation}
    P_x^3 = \frac{1-c}{v} \theta_x + \frac{1}{v} (1-\theta_x)
     = -\frac{c}{v} \theta_x + \frac{1}{v}.
\end{equation}
Thus, we have
\begin{equation}
    \mathbb{E}[\eta_x(W)] = \sum\limits_{i=1}^3 \lambda_i P_x^i = \frac{c}{v-c} \theta_x + \frac{v-2c}{v(v-c)}.
\end{equation}
\end{IEEEproof}

\bibliographystyle{IEEEtran}
\bibliography{ref}

\newpage
\setcounter{page}{1}
\section*{Supplementary Material}
Here, we consider four $(P_U,\tilde{Q})$'s constructed in Section~\ref{sec:opt_mech}, and the estimator $\tilde{\theta}_n$ in~\eqref{eq:opt_est}.
Also, we use the same $\lambda_i$'s and $P_x^i$'s in Appendices~\ref{app:lem_1},~\ref{app:lem_2}, or~\ref{app:lem_3}.

    \subsection{Proof of Lemma~\ref{lem:var_eta_conc}}
\begin{IEEEproof}
\subsubsection{Case 1} Let $(P_U,\tilde{Q})$ be the resolution of $Q$, which is in Case 1 of Section~\ref{sec:opt_mech}.
From~\eqref{eq:var_eta_x} and Appendix~\ref{app:lem_1},
\begin{align}
    \mathrm{Var}(\eta_x(W))
    = \lambda_1^2 P_x^1(1-P_x^1) &+ \lambda_2^2 P_x^2(1-P_x^2) \nonumber
    \\&- 2 \lambda_1 \lambda_2 P_x^1 P_x^2.
\end{align}
Because $P_x^1$ and $P_x^2$ are linear functions of $\theta_x$, $\mathrm{Var}(\eta_x(W))$ is clearly a second order polynomial in $\theta_x$.
Now, we focus on the coefficient of $\theta_x^2$ of such polynomial, denoted by $S$.
Let $\rho = \frac{v(c-d)}{2(v-1)(c+d)}$, the coefficient of $\theta_x$ in $P_x^1$.
Then,
\begin{equation}
    S/\rho^2 = -(\lambda_1^2 + \lambda_2^2) +2 \lambda_1 \lambda_2  = -(\lambda_1 - \lambda_2)^2  < 0.
\end{equation}

\subsubsection{Case 2}
Let $(P_U,\tilde{Q})$ be the resolution of $Q$, which are in Case 2 of Section~\ref{sec:opt_mech}.
Similar to 1), Appendix~\ref{app:lem_2} shows that $P_x^i$'s are linear functions of $\theta_x$, and~\eqref{eq:var_eta_x} implies that $\mathrm{Var}(\eta_x(W))$ is a second order polynomial in $\theta_x$.
Let $S$ be the coefficient of $\theta_x^2$ of such polynomial.
Note that for all $i \in [4]$, the coefficient of $\theta_x$ of $P_x^i$ is $\pm \rho$, $\rho = \frac{(\alpha+1)(c-d)}{2(2\alpha+1)(c+d)}$.
Thus, we have
\begin{align}
    S/\rho^2 & = -\sum\limits_{i=1}^4 \lambda_i^2
    \\&\quad- 2 \left( \lambda_1 \lambda_2 - \lambda_1 \lambda_3 - \lambda_1 \lambda_4 - \lambda_2 \lambda_3 - \lambda_2 \lambda_4 + \lambda_3 \lambda_4 \right) \nonumber
    \\& = - (\lambda_1 - \lambda_3 + \lambda_2 - \lambda_4)^2 < 0.
\end{align}

\subsubsection{Case 3, 4}
Let $(P_U,\tilde{Q})$ and $Q$ be the privacy mechanism with shared randomness and the stochastic matrix constructed in Case 3 or 4 in Section~\ref{sec:opt_mech}.
Similar to 1), Appendix~\ref{app:lem_3} shows that $P_x^i$'s are linear functions of $\theta_x$, and~\eqref{eq:var_eta_x} implies that $\mathrm{Var}(\eta_x(W))$ is a second order polynomial in $\theta_x$.
Let $S$ be the coefficient of $\theta_x^2$ of such polynomial.
Note that for all $i \in [3]$, the coefficient of $\theta_x$ of $P_x^i$ is $\pm \rho$, $\rho = c/v$ ($c=\delta$ for Case 3 and $c=e^\gamma-1$ for Case 4).
Thus, we have
\begin{align}
    S/\rho^2 &= - (\lambda_1^2 + \lambda_2^2 + \lambda_3^2) -2 ( \lambda_1 \lambda_2 - \lambda_1 \lambda_3 - \lambda_2 \lambda_3)
     \\ &= - (\lambda_1 - \lambda_3 + \lambda_2)^2 < 0.
\end{align}
\end{IEEEproof}

\subsection{Proof of Proposition~\ref{prop:PUT_SR_achiev}}
\begin{IEEEproof}
As we mentioned in Section~\ref{sec:Error_opt_scheme}, we first calculate $P_x^i$'s for $\theta = \mathbf{1}/v$, and substitute them into~\eqref{eq:var_eta_x}.
We calculate the worst-case MSEs of the optimal schemes with shared randomness in four cases, and each of them corresponds to each cases in Section~\ref{sec:opt_mech}.

\subsubsection{Case 1}
We denote $P_x^i$'s and $\lambda_i$'s as in Appendix~\ref{app:lem_1}.
For $\theta = \mathbf{1}/v$, $P_x^1 = 1/2$.
Thus,~\eqref{eq:var_eta_x} gives
\begin{equation}
    \mathrm{Var}(\eta_x(W)) = \frac{1}{4}(\lambda_1 - \lambda_2)^2 = \frac{(c-d)^2}{v^2(c+d)^2}. 
\end{equation}
Then, by~\eqref{eq:eta_first_coeff_1},~\eqref{eq:MSE_concave} and~\eqref{eq:MSE_unif}, we have
\begin{align}
    \!\!\! R_{n,v}(P_U,\tilde{Q},\tilde{\theta}_n) \!&=\! v \mathrm{Var}(\eta_x(W)) \left(\frac{(v-1) (c+d)^2}{(c-d)^2}\right)^2
    \\ \!&=\! \frac{(v-1)^2(c+d)^2}{v(c-d)^2}.
\end{align}
Plugging $(c,d) = \left( \frac{e^\epsilon+\delta}{e^\epsilon+1}, \frac{1-\delta}{e^\epsilon+1} \right)$ gives the desired result.

\subsubsection{Case 2}
We denote $P_x^i$'s and $\lambda_i$'s as in Appendix~\ref{app:lem_2}, and $v = 2\alpha + 1$.
For $\theta = \mathbf{1}/v$,
\begin{align}
    P_x^1 &= \frac{\alpha(\alpha c + (\alpha+1)d)}{(2\alpha+1)^2 (c+d)},
    \\ P_x^2 &= \frac{(\alpha+1)((\alpha+1) c + \alpha d)}{(2\alpha+1)^2 (c+d)},
    \\ P_x^3 &= \frac{(\alpha+1)(\alpha c + (\alpha+1) d)}{(2\alpha+1)^2 (c+d)},
    \\ P_x^4 &= \frac{\alpha((\alpha+1) c + \alpha d)}{(2\alpha+1)^2 (c+d)}.
\end{align}
Thus,~\eqref{eq:var_eta_x} gives
\begin{equation}
    \mathrm{Var}(\eta_x(W)) = \frac{(c-d)^2}{(2\alpha+1)^2 \left( c + \frac{\alpha}{\alpha + 1} d \right)\left( c + \frac{\alpha+1}{\alpha} d \right)}.
\end{equation}
By~\eqref{eq:eta_first_coeff_2},~\eqref{eq:MSE_concave} and~\eqref{eq:MSE_unif}, we have
\begin{align}
    &R_{n,v}(P_U,\tilde{Q},\tilde{\theta}_n) \nonumber
    \\ &= v \mathrm{Var}(\eta_x(W)) \left(\frac{2\alpha\left( c + \frac{\alpha}{\alpha + 1} d \right) \left( c + \frac{\alpha+1}{\alpha} d \right)}{(c-d)^2}\right)^2
    \\& = \frac{(v-1)^2}{v} \cdot \frac{ \left( c + \frac{\alpha}{\alpha + 1} d \right) \left( c + \frac{\alpha+1}{\alpha} d \right)}{(c-d)^2}.
\end{align}
Plugging $(c,d) = \left(\frac{e^\epsilon+\delta}{e^\epsilon+1}, \frac{1-\delta}{e^\epsilon+1} \right)$ gives the desired result.

\subsubsection{Case 3, 4}
We denote $P_i$'s and $\lambda_i$'s as in Appendix~\ref{app:lem_3}.
For $\theta = \mathbf{1}/v$,
\begin{equation}
    P_x^1 = \frac{c}{v^2}, \; P_x^2 = \frac{(v-c)(v-1)}{v^2},  \; P_x^3 = \frac{v-c}{v^2}.
\end{equation}
Thus,~\eqref{eq:var_eta_x} gives
\begin{equation}
    \mathrm{Var}(\eta_x(W)) = \frac{c(v-1)}{v^2(v-c)}.
\end{equation}
By~\eqref{eq:eta_first_coeff_3},~\eqref{eq:MSE_concave} and~\eqref{eq:MSE_unif}, we have
\begin{align}
    R_{n,v}(P_U,\tilde{Q},\tilde{\theta}_n) &= v \mathrm{Var}(\eta_x(W))  \left( \frac{v-c}{c} \right)^2 \\
    &= \frac{(v-1)(v-c)}{vc}.
\end{align}
Plugging $c = \delta$ and $e^\gamma - 1$ yield the desired results for cases 3 and 4, respectively.
\end{IEEEproof}

\end{document}